\documentclass[journal=jctcce,manuscript=article,layout=twocolumn]{achemso}
\setkeys{acs}{articletitle = true, chaptertitle = true, doi = true}

\usepackage[version=3]{mhchem} 
\usepackage[T1]{fontenc}       
\usepackage{mathptmx}

\makeatletter
\let\l@addto@macro\relax
\makeatother
\usepackage[fontsize=9pt]{scrextend}

\usepackage{graphicx}  
\usepackage{bm}        
\usepackage{bbm}       
\usepackage{amssymb}   
\usepackage{amsmath}   
\usepackage{physics}   
\usepackage{textgreek} 
\usepackage{subfigure} 
\usepackage{multirow}  
\usepackage{booktabs}  
\usepackage{hyperref}  
\hypersetup{colorlinks=true, linkcolor=black, urlcolor=black, citecolor=black,
plainpages=false, bookmarksnumbered=true}
\usepackage{xurl}      
\usepackage{siunitx}   
\usepackage{xcolor}    
\usepackage[utf8]{inputenc} 
\newcommand{\ve}[1]{\bm{#1}}

\author{Yannick J. Franzke}
\affiliation{Institute of Nanotechnology, Karlsruhe
Institute of Technology (KIT), Kaiserstra\ss{}e 12, 76131 Karlsruhe, Germany}
\email{yannick.franzke@kit.edu}

\author{Christof Holzer}
\affiliation{Institute of Quantum Materials and Technologies, Karlsruhe
Institute of Technology (KIT), Kaiserstra\ss{}e 12, 76131 Karlsruhe,
Germany}
\email{christof.holzer@kit.edu}

\title{Application of SAP-X2C to Spectroscopy and Comparison to Screened Nuclear Spin--Orbit Approximations}

\begin{document}



\begin{abstract}
Recently, Surjuse and Valeev [\textit{J. Chem. Theory Comput.} \textbf{22},
3443--3452 (2026).] suggested to use a simple superposition of
atomic potentials to account for the two-electron picture-change
error in one-electron exact two-component (SAP-X2C) theory. 
Herein, we generalize this ansatz to analytical derivative theory
and apply SAP-X2C to molecular spectroscopy.
The accuracy is assessed for NMR, EPR, M\"ossbauer, UV/vis, as
well as X-ray absorption spectroscopy. A thorough comparison
with four-component results and the even simpler screened nuclear
spin--orbit (SNSO) approximation reveals that both SNSO-X2C and
SAP-X2C perform excellently for spectroscopic properties, with
SAP-X2C yielding slightly lower errors. Another major advantage is
its well defined thermodynamic limit and less empirical nature.
Therefore, SAP-X2C may be expected to become the default choice
to mitigate the two-electron picture-change error in density
functional theory approaches for spectroscopy thanks to its
accuracy, simplicity, and efficiency.
More complicated approaches to account for this error based on
atomic mean-field ans\"atze may still be relevant for high-level
correlated methods and highly accurate thermochemistry.
\end{abstract}

\section{Introduction}
\label{sec:introduction}
The consideration of special relativity is a key component of many
quantum-chemical program suites in order to support accurate computational
methods throughout the complete periodic table of elements.
\cite{Pyykko:Physics.2012, Pyykko:Relativistic.2012}
Relativistic effects are accounted for with either effective core potentials
\cite{Dolg.Cao:Relativistic.2012} or all-electron approaches.
\cite{Liu:Ideas.2010, Peng.Reiher:Exact.2012, Saue:Relativistic.2011} 
A key advantage of all-electron relativistic approaches over the
computationally cheaper relativistic pseudopotentials is their
applicability to study nuclear magnetic resonance (NMR) and
electron paramagnetic resonance (EPR) spectroscopy of compounds or
materials with heavy elements. \cite{Kaupp.Buhl.ea:Calculation.2004,
Autschbach:Perspective.2012, Autschbach:Relativistic.2014,
Reiher.Wolf:2015, InC-Repisky.Komorovsky.ea:2016, InB-Neese:2017,
Liu:2017, InC-Komorovsky:2024, Aucar:2025}
Magnetic resonance spectroscopy is of great importance for
the characterization of novel compounds \cite{Vaara:Theory.2007}
or the developments of new quantum technologies based on single molecule magnets
\cite{Wolfowicz.Tyryshkin.ea:Atomic.2013, Zadrozny.Niklas.ea:Millisecond.2015, 
Shiddiq.Komijani.ea:Enhancing.2016, Gaita-Arino.Luis.ea:Molecular.2019,
Kundu.White.ea:Clock.2022,Chilton:Molecular.2022, Smith.Hruby.ea:Identification.2024,
Ngo.McClain.ea:Large.2025, Liu.Chen.ea:linear.2025, Moreno-Pineda.Wernsdorfer:Magnetic.2025}
and the availability of accurate relativistic all-electron approaches
is crucial for \textit{in-silico} studies.

Overall, relativistic all-electron exact two-component (X2C) theory
\cite{Dyall:Interfacing.1997, Dyall:Interfacing.1998, Dyall.Enevoldsen:Interfacing.1999,
Dyall:Interfacing.2001, Kutzelnigg.Liu:Quasirelativistic.2005,
Liu.Kutzelnigg:Quasirelativistic.2007, Liu.Peng:Infinite-order.2006,
Liu.Peng:Erratum.2006, Ilias.Saue:infinite-order.2007, Liu.Peng:Exact.2009,
Peng.Liu.ea:Making.2007} has become the \textit{de-facto} standard in
most modern quantum chemical codes. 
\cite{Liu:Essentials.2020, Sherrill.Manolopoulos.ea:Electronic.2020,
Gagliardi:Sustainable.2023, Crawford.Dreuw:Quantum.2026}
In X2C, the negative energy subspace is decoupled by a simple diagonalization
to arrive at an ``electrons-only'' Hamiltonian.
\cite{Liu:Ideas.2010, Peng.Reiher:Exact.2012, Saue:Relativistic.2011}
For practical simplicity, this decoupling is often performed
for the one-electron Hamiltonian only and the two-electron infrastructure
is applied without any modifications. This one-electron X2C (1e-X2C)
ansatz introduces the so-called two-electron picture-change error (2ePCE)
and leads to different results from 1e-X2C and the parent four-component
DIRAC calculations.

To mitigate the two-electron picture-change error, Boettger introduced a simple
rescaling approach of the spin--orbit contributions to the Hamiltonian, which was
termed screened nuclear spin--orbit (SNSO) approximation. \cite{Boettger:Approximate.2000}
This accounts for the screening of the nuclear charge due to the
electron--electron interaction  \cite{Blume.Watson:Theory.1962, Blume.Watons:Theory.1963}
and was successfully applied for spin--orbit
splittings in low-order Douglas--Kroll--Hess (DKH) theory.
\cite{Hess:Relativistic.1986, Jansen.Hess:Revision.1989}
The original parameters were directly set based on the orbital shell of
the basis functions. Later, the corresponding screening parameters were
refined for X2C calculations by the groups of Cremer
\cite{Filatov.Zou.ea:Spin-orbit.2013, Zou.Filatov.ea:Analytical.2015,
Yoshizawa.Zou.ea:Calculations.2016, Yoshizawa.Filatov.ea:Calculation.2019}
and Li. \cite{Ehrman.Martinez-Baez.ea:Improving.2023}
As suggested by its name, SNSO only accounts for the spin--orbit two-electron
interaction but not the scalar-relativistic 2ePCE. Nevertheless,
the SNSO approximation was already successfully applied for a broad range
of spectroscopic properties and was shown to significantly reduce the
deviations of the two-component (2c) Hamiltonian to the four-component (4c) reference.
\cite{Yoshizawa.Hada:Calculations.2017, Franzke.Mack.ea:NMR.2021, Franzke.Holzer:Exact.2023,
Franzke.Yu:Hyperfine.2022, Franzke.Yu:Quasi-Relativistic.2022, 
Yoshizawa.Zou.ea:Calculations.2016, Kehry.Franzke.ea:Quasirelativistic.2020,
Yoshizawa.Filatov.ea:Calculation.2019, Ehrman.Martinez-Baez.ea:Improving.2023}
Therefore, SNSO-X2C is a major improvement upon 1e-X2C.
This is especially true for low-scaling approaches such as
Hartree--Fock (HF) and density functional theory (DFT), where
a cost-efficient approach to consider the two-electron picture-change
error is pivotal. \cite{Saue:Relativistic.2011}

More complicated approaches based on molecular mean-field 
\cite{Sikkema.Visscher.ea:molecular.2009} (mmf) or
atomic mean-field (amf) corrections \cite{Wodynski.Kaupp:Density.2019,
Knecht.Repisky.ea:Exact.2022, Repisky.Komorovsky.ea:X2C.2025,
Liu.Cheng:atomic.2018, Zhang.Cheng:Atomic.2022,
Wang.Zhang.ea:Relativistic.2025, Wang.Wang.ea:Relativistic.2026}
such as amfX2C \cite{Knecht.Repisky.ea:Exact.2022}
and its extended version eamfX2C \cite{Knecht.Repisky.ea:Exact.2022}
were suggested in the literature for X2C.
These allow for an accurate consideration of the 2ePCE for total
energies and spin--orbit splittings at a reduced cost compared to
four-component calculations.
Yet, these ans\"atze are still far more complex than SNSO treatments.
Moreover, amfX2C and related approaches were not yet applied to
higher-order derivatives such as those needed for NMR spectra.

Recently, \cite{Surjuse.Valeev:SAP-X2C.2026}
Surjuse and Valeev suggested to use a superposition of
atomic potentials (SAP) to account for the two-electron interaction in the
relativistic decoupling. Here, an SAP contribution
\cite{Lehtola:Assessment.2019, Lehtola.Visscher.ea:Efficient.2020}
is added to the bare one-electron potential terms and then the X2C diagonalization
step is performed. Subsequently, the non-relativistic SAP integrals
are subtracted to avoid the need for modifications of the actual
two-electron infrastructure. Notably, this ansatz results in a well
defined thermodynamic limit. Compared to the amfX2C approaches,
this is a much simpler scheme and the authors have shown that SAP-X2C
rather accurately recaptures the $Z$-adjusted total energies, spin--orbit
splittings, and geometries of diatomic systems. SAP-X2C is further
closely related to a reformulation of the model potential approach
\cite{Wullen:Molecular.1998} for two-electron picture-change errors
in DKH theory proposed by van W\"ullen and Michauk in 2005.
\cite{Wullen.Michauk:Accurate.2005}
Their scheme uses Gaussian $s$-functions for the 4c model potential
in the decoupling step and a 2c model density in the subtraction afterwards.
However, the basis sets are not deposited at a public repository, whereas
the publicly available basis set of Lehtola \textit{et al.} 
\cite{Lehtola.Visscher.ea:Efficient.2020} is used for SAP-X2C.
This facilitates the incorporation of SAP-X2C in almost all relativistic
quantum-chemistry codes. However, its application and accuracy for
spectroscopy---especially magnetic resonance spectroscopy remains to be
demonstrated.

In the present work, we present the straightforward and simple
generalization of SAP-X2C to molecular spectroscopy and assess its
accuracy compared to the already established SNSO approximations for
NMR, EPR, and M\"ossbauer properties as well as UV/vis and X-ray
absorption spectra.

\section{Theory \& Implementation}
\label{sec:theory}

\subsection{SAP-X2C Theory}
\label{subsec:theo-sap}
The one-electron X2C (1e-X2C) Hamiltonian reads
\begin{equation}
    \textbf{h}^{\text{1e-X2C}} = \textbf{R}^{\dag} \left( \textbf{V} + \textbf{T} \textbf{X} + \textbf{X}^{\dag} \textbf{T} + \textbf{X}^{\dag} \left[ \frac{\textbf{W}}{4 c^2} - \textbf{T} \right] \textbf{X} \right) \textbf{R}
\end{equation}
with the decoupling matrix \textbf{X} and the renormalization matrix \textbf{R} following from the
solutions of the one-electron Dirac equation in a restricted kinetic balance.
\cite{Kutzelnigg.Liu:Quasirelativistic.2006, Liu.Peng:Exact.2009, Ilias.Saue:infinite-order.2007,
Peng.Middendorf.ea:efficient.2013} As usual, $c$ refers to the speed of light.
All other ingredients are one-electron integrals, i.e.\ the kinetic energy matrix \textbf{T},
the scalar potential \textbf{V}, and the relativistically modified potential \textbf{W}.
Only the latter is not block-diagonal in the 2c basis space but can be partitioned into a
spin-independent scalar and a spin-dependent spin--orbit integral contribution according to
\begin{equation}
\begin{split}
    W_{\mu \nu} = & \mel{\mu}{\big(\boldsymbol{\sigma} \cdot  \textbf{p} \big) V \big(\boldsymbol{\sigma} \cdot \textbf{p} \big)}{\nu} =
    W_{\mu \nu}^{\text{SR}} + W_{\mu \nu}^{\text{SO}} \\ = &  \textbf{1}_2 \mel{\mu}{\textbf{p} V \cdot \textbf{p}}{\nu}
    + \text{i} \boldsymbol{\sigma} \cdot \mel{\mu}{\textbf{p} V \crossproduct \textbf{p}}{\nu}
\end{split}
\end{equation}
with the atom-centered basis functions $\mu, \nu$ and the vector $\boldsymbol{\sigma}$ consisting
of the Pauli spin matrices. The X2C decoupling transformation is carried out in the uncontracted
basis set and the contraction is performed afterwards to use standard electronic basis sets.

In SAP-X2C, \cite{Surjuse.Valeev:SAP-X2C.2026} the potential terms \textbf{V} and \textbf{W} account
for the missing two-electron integrals in an approximate fashion by introducing an SAP basis set
consisting of contracted and atom-centered $s$-functions given by
\begin{equation}
     \alpha_N (\textbf{r}) = \sum_p c_{N, p} \left( \frac{\pi}{\zeta_{N, p}} \right)^{3/2} \exp\left( - \zeta_{N, p} \left|\textbf{r} - \textbf{R}_N \right|^2 \right)
\end{equation}
Here, $\zeta_{N, p}$ are the SAP exponents and $\textbf{R}_N$ denotes the position of nucleus $N$.
The coefficients $c_p$ are normalized to the absolute charge of the nucleus,
i.e.\ $\sum_p c_{N, p} = - Z_N$, leading to
\begin{eqnarray}
  V_{\mu \nu}^{\text{SAP}} & = & {\phantom{-}} V_{\mu \nu} + V_{\mu \nu}^{\text{e}} \\
  V_{\mu \nu}^{\text{e}} & = & - \sum_N \left(\alpha_{N} | \mu \nu \right) \\
  W_{\mu \nu}^{\text{SAP}} & = & {\phantom{-}} W_{\mu \nu} + W_{\mu \nu}^{\text{e}} \\
  W_{\mu \nu}^{\text{e}} & = & - \sum_N \left(\alpha_{N} | \left[ \left(\boldsymbol{\sigma} \cdot \textbf{p} \right) \mu \right] \left[\left(\boldsymbol{\sigma} \cdot \textbf{p} \right) \nu \right] \right)
\end{eqnarray}
We applied the Mulliken notation to indicate the close relationship to three-center-two-electron integrals,
\begin{equation}
    \left(\alpha_{N} | \mu \nu \right) = \int \mathrm{d}\textbf{r} \, \mathrm{d}\textbf{r'} \, \alpha_N (\textbf{r}) \frac{1}{| \textbf{r} - \textbf{r'}|} \mu (\textbf{r'}) \nu (\textbf{r'})
\end{equation}
This finally results in the SAP-X2C Hamiltonian
\begin{equation}
\begin{split}
    \textbf{h}^{\text{SAP-X2C}} = \textbf{R}^{\dag} \left( \textbf{V}^{\text{SAP}} + \textbf{T} \textbf{X} + \textbf{X}^{\dag} \textbf{T} + \textbf{X}^{\dag} \left[ \frac{\textbf{W}^{\text{SAP}}}{4 c^2} - \textbf{T} \right] \textbf{X} \right) \textbf{R} - \textbf{V}^{\text{e}},
    \end{split}
\end{equation}
where the last term counteracts the two-electron potential integrals after the X2C decoupling
step. Application of the product rule to form analytical derivatives for molecular
properties \cite{Cheng.Gauss:Analytic.2011, Zou.Filatov.ea:Development.2012,
Cremer.Zou.ea:Dirac-exact.2014, Cheng.Stopkowicz.ea:Analytic.2014, InB-Franzke:2025}
and spectroscopy is straightforward. The same holds for the magnetic balance condition.
\cite{Komorovsky.Repisky.ea:fully.2008} Importantly, SAP-X2C accounts for both
scalar-relativistic and spin--orbit two-electron picture-change errors.

\subsection{Implementation of SAP-X2C for Properties}
\label{subsec:sap-implementation}
The SAP integrals can be easily evaluated based on the one-electron
integral routines for the finite nucleus model. Here, only the charge
is replaced with the coefficient and an additional loop over the different
exponents for each atom is required.
This simple implementation allowed us to directly extend the SAP-X2C scheme
to NMR properties, \cite{Franzke.Weigend:NMR.2019, Franzke.Holzer:Exact.2023,
Franzke:Reducing.2023, Franzke.Mack.ea:NMR.2021}
EPR properties, \cite{Franzke.Yu:Hyperfine.2022, Franzke.Yu:Quasi-Relativistic.2022, 
Gillhuber.Franzke.ea:Efficient.2021, Bruder.Franzke.ea:Paramagnetic.2022,
Bruder.Franzke.ea:Zero-Field.2023}
M\"ossbauer properties, \cite{Holzer.Franzke:General.2025}
and analytical gradients \cite{Franzke.Middendorf.ea:Efficient.2018,
Franzke.Spiske.ea:Segmented.2020} based on the existing infrastructure
in TURBOMOLE. \cite{Franzke.Holzer.ea:TURBOMOLE.2023, TURBOMOLE}
Likewise, excited state properties can be extended.
\cite{Kehry.Franzke.ea:Quasirelativistic.2020, Himmelsbach.Holzer:Excited.2024}
These are mostly affected through the change in the ground-state density
upon the consideration of the two-electron picture-change correction in X2C,
as the impact of the relativistic picture-change transformation to the
transition dipole operators is often small. In the course of this work,
we implemented the Kramers-unrestricted non-collinear exchange--correlation
kernel of the Scalmani--Frisch formalism in its original
\cite{Scalmani.Frisch:New.2012, Egidi.Sun.ea:Two-Component.2017}
and its regularized version \cite{Komorovsky.Cherry.ea:Four-component.2019}
up to current-dependent \textit{meta}-generalized gradient approximations
with range-separated hybrid and local hybrid functionals into TURBOMOLE,
\cite{Balasubramani.Chen.ea:TURBOMOLE.2020, Franzke.Holzer.ea:TURBOMOLE.2023, TURBOMOLE}
see the Appendix.

An important advantage of SAP-X2C is that it straightforwardly enables
periodic calculations due to charge neutrality and a well defined thermodynamic
limit. \cite{Surjuse.Valeev:SAP-X2C.2026} However, the SAP basis sets of
ref~\citenum{Lehtola.Visscher.ea:Efficient.2020} include very diffuse basis
functions which is detrimental for periodic systems and near-field/far-field
partitioning schemes. \cite{Challacombe.White.ea:Periodic.1997, Kudin.Scuseria:Revisiting.2004,
Burow.Sierka.ea:Resolution.2009, Kadek.Repisky.ea:All-electron.2019, Sharma.Franzke.ea:Density.2025}
Therefore, we refitted the large relativistic SAP basis set of this
reference under three constraints: charge neutrality, conserved first and
third radial moments. These new basis sets are intended for SAP-X2C of periodic
systems and are given in the Supporting Dataset together with their assessment.
Unless explicitly stated differently, we employ the large GRASP-optimized SAP
basis. \cite{Lehtola.Visscher.ea:Efficient.2020}

\subsection{SNSO-X2C Theory}
\label{subsec:theo-snso}
In the SNSO approximations, the spin-dependent contribution of the
relativistically modified potential matrix (SNSO(W)) or the Hamiltonian
matrix (SNSO(Ham)) is rescaled according to
\begin{equation}
    W_{\mu \nu}^{\text{SO}} = \left(1 - \sqrt{\frac{Q(l_{\mu})}{Z_{\mu}}}
    \sqrt{\frac{Q(l_{\nu})}{Z_{\nu}}}\right) W_{\mu \nu}^{\text{SO}}
\end{equation}
depending on the angular momentum $l$ of the basis functions and the
charge $Z$ of their atom centers. Especially for analytical derivatives,
rescaling $\textbf{W}^{\text{SO}}$ instead of the Hamiltonian matrix $\textbf{H}^{\text{SO}}$
is the preferred choice in terms of simplicity \cite{Wullen.Michauk:Accurate.2005,
Zou.Filatov.ea:Analytical.2015} and its implementation in analytical derivative
theory or for the picture-change correction of expectation values is trivial.
Therefore, we will only show results with this SNSO(W) ansatz herein.

Currently, three different sets of SNSO parameters $Q$ are available in the literature.
\begin{enumerate}
    \item The original ones by Boettger for low-order DKH theory
    given by \cite{Boettger:Approximate.2000}
    \begin{eqnarray}
    Q(1) & = & 2 \\
    Q(2) & = & 10 \\
    Q(3) & = & 28
    \end{eqnarray}
    \item A modified version reoptimized for X2C by Filatov, Zou, and Cremer
    denoted mSNSO with \cite{Filatov.Zou.ea:Spin-orbit.2013,
    Yoshizawa.Zou.ea:Calculations.2016, Yoshizawa.Filatov.ea:Calculation.2019}
    \begin{eqnarray}
    Q(1) & = & 2.34 ~ \text{erf} \left(34500/ \zeta_p^2 \right) \\
    Q(2) & = & 11.00 \\
    Q(3) & = & 28.84
    \end{eqnarray}
    \item The Dirac--Coulomb (SNSO-DC) as well as the (row-dependent) Dirac--Coulomb--Breit (SNSO-DCB) 
    schemes of the Li group with the SNSO-DC parameters reading \cite{Ehrman.Martinez-Baez.ea:Improving.2023}
    \begin{eqnarray}
    Q'(1) & = & 2.32 \\
    Q'(2) & = & 10.64 \\
    Q'(3) & = & 28.38
    \end{eqnarray}
\end{enumerate}
SNSO, SNSO-DC, and mSNSO treat the high angular-momentum basis functions the same way as
\begin{equation}
Q(l \geq 4)  =  l (l + 1)(2l + 1)/3 = 60, 110, \dots
\end{equation}
and naturally no correction is applied for the $s$-type functions.
Among these schemes, the mSNSO approach further reassigns the angular
momentum numbers to improve the virtual spinor states according to
\begin{equation}
Q'(l) = 
\begin{cases}
Q(l) &\mbox{if } Z > Q(l) \\
Q(l') & \mbox{if } Z \leq Q(l)
\end{cases}
\end{equation}
with $l'$ denoting the maximal orbital angular momentum number so that $Z > Q(l')$.
\cite{Filatov.Zou.ea:Spin-orbit.2013}

The SNSO approximation with various parameters is already available
in TURBOMOLE \cite{Balasubramani.Chen.ea:TURBOMOLE.2020, Franzke.Holzer.ea:TURBOMOLE.2023, TURBOMOLE}
or COLOGNE \cite{Cremer.Zou.ea:Dirac-exact.2014, Yoshizawa.Hada:Calculations.2017,
Yoshizawa.Filatov.ea:Calculation.2019, Cologne},
for all X2C modules therein and a broad range of molecular properties.

\section{Computational Methods}
\label{sec:methods}
First, the total self-consistent field (SCF) energy and spinor eigenvalues
of Au$_2$ (bond distance $4.67$\,a.u.) were calculated with SAP-X2C
\cite{Surjuse.Valeev:SAP-X2C.2026} and the three SNSO approaches listed
above \cite{Boettger:Approximate.2000, Ehrman.Martinez-Baez.ea:Improving.2023,
Filatov.Zou.ea:Spin-orbit.2013, Yoshizawa.Zou.ea:Calculations.2016, Yoshizawa.Filatov.ea:Calculation.2019}
as well as 1e-X2C \cite{Peng.Middendorf.ea:efficient.2013} at the Hartree--Fock
(HF) level. The SCF energy of an isolated Au atom was calculated accordingly.
Results are compared to the amfX2C and four-component Dirac--Coulomb--Hartree--Fock (DHF)
approach of ref~\citenum{Repisky.Komorovsky.ea:X2C.2025}.
Therefore, the uncontracted Dyall-VTZ basis set \cite{Dyall:Relativistic.2004, Dyall-Repo}
was applied in the point charge model. SCF energies were converged up
to $10^{-10}$\,$E_{\text{h}}$ and a criterion of $10^{-9}$
for the root-mean square of the density matrix change.
All 1e/SNSO/SAP-X2C calculations herein were performed with TURBOMOLE.
\cite{Franzke.Holzer.ea:TURBOMOLE.2023, TURBOMOLE} The resolution of
the identity approximations (RI-J, RI-K) were only applied if explicitly
stated. Otherwise, analytical Coulomb and exchange integrals were computed.
The 4c-DHF energy of an isolated Au atom was calculated with ReSpect herein to
compute the atomization energy. \cite{Repisky.Komorovsky.ea:ReSpect.2020, Respect}

Second, the impact on the contact density for M{\"o}ssbauer spectroscopy
is studied for Hg, HgF, HgF$_2$, and HgF$_4$ with respect to four-component
DHF and Dirac--Coulomb--Kohn--Sham (DKS) values. \cite{Knecht.Fux.ea:Mossbauer.2011}
For consistency, we used the finite nucleus model \cite{Visscher.Dyall:Dirac-Fock.1997}
and the modified Dyall-CVQZ+2s+1p basis set for Hg \cite{Dyall:Relativistic.2004,
Dyall.Gomes:Revised.2009, Dyall-Repo, Knecht.Fux.ea:Mossbauer.2011} and
the decontracted aug-cc-pVQZ basis for F. \cite{Dunning:Gaussian.1989}
Additionally, 2c DFT calculations with the PBE \cite{Perdew.Burke.ea:Generalized.1996}
functional using very large grids \cite{Franzke.Tress.ea:Error-consistent.2019}
(grid size 5a) and the modified Dyall-CVTZ+2s+1p/aug-cc-pVTZ basis set
\cite{Dyall:Relativistic.2004, Dyall.Gomes:Revised.2009, Dyall-Repo,
Knecht.Fux.ea:Mossbauer.2011, Dunning:Gaussian.1989}
in decontracted form were carried out.
Structure parameters were taken from ref~\citenum{Knecht.Fux.ea:Mossbauer.2011}
and SCF energies were converged up to $10^{-9}$\,$E_{\text{h}}$.
Open-shell X2C-DFT calculations of HgF used the canonical 2c formalism.
\cite{Armbruster.Weigend.ea:Self-consistent.2008, Baldes.Weigend:Efficient.2013}

Third, NMR isotropic shielding constants of atoms and molecules
were calculated with two-component HF \cite{Franzke.Holzer:Exact.2023}
and four-component DHF theory \cite{Komorovsky.Repisky.ea:fully.2008, 
Komorovsky.Repisky.ea:Fully.2010, Novotny.Vicha.ea:Linking.2017} in the point-charge model.
The magnetic balance condition and gauge-including atomic orbitals
were employed throughout for consistency. For the atomic calculations, uncontracted
Dyall-VDZ basis sets were applied. \cite{Dyall:Relativistic.2004, Dyall:Relativistic.2006,
Dyall:Relativistic.2009, Dyall:Relativistic.2016, Dyall-Repo}
Molecular calculations used the uncontracted Dyall-VDZ basis set for Br, I, At, Se, Te,
Po, As, Sb, Bi, Ge, Sn, and Pb, \cite{Dyall:Relativistic.2006, Dyall-Repo}
whereas the decontracted cc-pVDZ basis set was employed for H, F, Cl, O, S, N, P, C, and Si.
\cite{Dunning:Gaussian.1989, Woon.Dunning:Gaussian.1993}
The SCF procedure was converged with up to at least $10^{-9}$\,$E_{\text{h}}$
and $10^{-9}$ for the root-mean square of the density matrix change.
Response equations are converged with a criterion of $10^{-7}$
for the norm of the residuum. \cite{Kehry.Franzke.ea:Quasirelativistic.2020}
Four-component Dirac--Coulomb calculations were performed accordingly with ReSpect.
\cite{Repisky.Komorovsky.ea:ReSpect.2020, Respect} Structure parameters are
listed in the Supporting Dataset.

Fourth, the accuracy of EPR properties is assessed with the 17 small transition-metal
complexes of ref~\citenum{Gohr.Hrobarik.ea:Four-Component.2015} using the
uncontracted Dyall-TZ basis set for the metal atom and Br, \cite{Dyall:Relativistic.2006, Dyall-Repo}
and the decontracted IGLO-III basis sets for the other atoms.
\cite{Kutzelnigg.Fleischer.ea:Deuterium.1991} The finite nucleus
model was considered for both the scalar potential and the vector
potential. \cite{Visscher.Dyall:Dirac-Fock.1997}
The canonical non-collinear formalism \cite{Armbruster.Weigend.ea:Self-consistent.2008, 
Baldes.Weigend:Efficient.2013} was applied for consistency with the four-component reference 
\cite{Gohr.Hrobarik.ea:Four-Component.2015} using the PBE0-40HF functional
\cite{Perdew.Burke.ea:Generalized.1996, Gohr.Hrobarik.ea:Four-Component.2015}
with very large grids \cite{Franzke.Tress.ea:Error-consistent.2019}
(grid size 5a) and the XCFun library. \cite{Ekstrom.Visscher.ea:Arbitrary-Order.2010, XCFun}
The SCF procedure was converged with a threshold of $10^{-9}$\,$E_{\text{h}}$
for the energy. We stress that our implementation fully includes the derivative
of the decoupling and the renormalization matrix.
\cite{Franzke.Yu:Hyperfine.2022, Franzke.Yu:Quasi-Relativistic.2022}
For consistency with the 4c Dirac--Coulomb reference, the kinetic balance condition
and a common gauge origin at the metal atom were used
(although GIAOs and magnetic balance are available in our implementation).
Structures were taken from ref~\citenum{Gohr.Hrobarik.ea:Four-Component.2015}.
The g-shift $\Delta g$ was obtained in parts per thousand (ppt) relative to
the g factor of the free electron. Furthermore, the EPR parameters of
NpF$_6$ are studied at the PBE0-40HF/Dyall-VTZ/IGLO-III level.
\cite{Perdew.Burke.ea:Generalized.1996, Gohr.Hrobarik.ea:Four-Component.2015,
Dyall:Relativistic.2007A, Dyall-Repo, Kutzelnigg.Fleischer.ea:Deuterium.1991}
Again, basis sets were fully decontracted.
Structural parameters were taken from ref~\citenum{Notter.Bolvin:Optical.2009},
whereas 4c as well as amfX2C/eamfX2C reference values are taken from
ref~\citenum{Repisky.Komorovsky.ea:X2C.2025}. Accordingly, the
regularized Scalmani--Frisch approach was applied and the vector potential
was included with the point charge model. Overall, all molecules in the
EPR studies herein feature a doublet ground state.

Fifth, excitation energies and the UV/vis spectra were computed for the
charge-neutral hydrindacene bismuthinidene of ref~\citenum{Pang.Nothling.ea:Synthesis.2023}.
Here, the non-collinear Scalmani--Firsch DFT and time-dependent DFT (TDDFT) approach
was applied in its regularized form \cite{Franzke.Schosser.ea:Efficient.2024,
Komorovsky.Cherry.ea:Four-component.2019} with the x2c-TZVPall-2c basis set
\cite{Pollak.Weigend:Segmented.2017} for Bi and the x2c-SVPall  basis
\cite{Pollak.Weigend:Segmented.2017} for H and C in the finite nucleus model.
\cite{Visscher.Dyall:Dirac-Fock.1997}
The B3LYP, \cite{Becke:Density-functional.1988, Lee.Yang.ea:Development.1988, Stephens.Devlin.ea:Ab.1994}
CAM-B3LYP, \cite{Yanai.Tew.ea:new.2004}
TPSS, \cite{Tao.Perdew.ea:Climbing.2003}
TPSSh, \cite{Staroverov.Scuseria.ea:Comparative.2003}
and CHYF-B95 \cite{Holzer.Franzke:General.2025}
density functional approximations were applied with medium-sized grids
(grid size 3a) \cite{Franzke.Tress.ea:Error-consistent.2019} and
the conductor-like screening model (COSMO). \cite{Pausch:Consistent.2024}
The CAM-B3LYP functional expressions were evaluated with
Libxc. \cite{Lehtola.Steigemann.ea:Recent.2018, LIBXC.2024} All functionals
except for B3LYP are only applied for the first 10 excitations to calculate the
zero-field splitting.
The resolution of the identity approximation was employed for the Coulomb
\cite{Armbruster.Weigend.ea:Self-consistent.2008} and the HF exchange
integrals \cite{Franzke.Holzer.ea:NMR.2022} of the ground and
first 100 excited states with tailored basis sets (see Supporting Dataset).
Calculations were facilitated with the diagonal local approximation to the unitary decoupling
transformation (DLU). \cite{Peng.Reiher:Local.2012, Peng.Middendorf.ea:efficient.2013}
SCF energies were converged up to $10^{-9}$\,$E_{\text{h}}$ and the density matrix
with a criterion of $10^{-6}$. TDDFT response equations were converged with a criterion
of $10^{-6}$ for the norm of the residuum. \cite{Kehry.Franzke.ea:Quasirelativistic.2020}
The molecular structure optimized with B3LYP is available in the Supporting Dataset.

\begin{table*}[t]
    \centering
    \caption{SCF total energies $E$ and spinor eigenvalues $\epsilon$ of Au$_2$
    at the Hartree--Fock level with the decontracted Dyall-VTZ basis within the
    point-charge model in atomic units.
    Structure and results with amfX2C and 4c-DHF are taken from 
    ref~\citenum{Repisky.Komorovsky.ea:X2C.2025}. eamfX2C and 4c-DHF
    are virtually indistinguishable and therefore eamfX2C is omitted in the table.
    $\Delta_{\text{X2C}} = (E_{\text{X2C}} -E_{\text{4c-DHF}})/\sum_N Z_N^2$ denotes
    the $Z$-adjusted errors in m$E_{\text{h}}/e^2$, as introduced in ref~\citenum{Surjuse.Valeev:SAP-X2C.2026}
    with $Z = 79$ for Au.}
    \label{tab:energies}
    \begin{tabular}{
    l
    S[table-format = -5.5]
    S[table-format = -5.5]
    S[table-format = -5.5]
    S[table-format = -5.5]
    S[table-format = -5.5]
    S[table-format = -5.5]
    S[table-format = -5.5]
    }
    \toprule
    & {\text{1e-X2C}} & {\text{SNSO-X2C}} & {\text{SNSO-DC-X2C}} & {\text{mSNSO-X2C}} & {\text{SAP-X2C}} & {\text{amfX2C}} & {\text{4c-DHF}} \\
    \midrule
    $E$                   & -38065.91245 & -38062.23002 & -38061.68451 & -38061.90619 & -38084.87087 & -38079.02163 & -38079.02175 \\
    $\Delta_{\text{X2C}}$ & +1.05 & +1.35 & +1.39 & +1.37 & -0.47 & +0.00 & {\text{none}} \\
    $\epsilon_{1s_{1/2}}$ & -2983.57320 & -2983.64665 & -2983.65720 & -2983.65124 & -2986.24534 & -2987.71436 & -2987.71442 \\
    $\epsilon_{2s_{1/2}}$ & -531.80918 & -531.83453 & -531.83786 & -531.83605 & -532.16206 & -532.42258 & -532.42259 \\
    $\epsilon_{2p_{1/2}}$ & -510.48161 & -508.93639 & -508.69088 & -508.72654 & -509.61262 & -509.28158 & -509.28157 \\
    $\epsilon_{2p_{3/2}}$ & -440.76497 & -441.24297 & -441.31845 & -441.31976 & -441.55652 & -441.69302 & -441.69305 \\
                          & -440.76459 & -441.24261 & -441.31808 & -441.31939 & -441.55614 & -441.69265 & -441.69265 \\
    \dots & \dots & \dots & \dots & \dots & \dots & \dots & \dots  \\
    $\epsilon_{75}$       & -0.41787   & -0.41850   & -0.41853 & -0.41854 & -0.41839 & -0.41821 & -0.41821 \\
    $\epsilon_{76}$       & -0.40693   & -0.41059   & -0.41083 & -0.41097 & -0.41090 & -0.41168 & -0.41170 \\
    $\epsilon_{77}$       & -0.38823   & -0.39139   & -0.39163 & -0.39175 & -0.39166 & -0.39196 & -0.39198 \\
    $\epsilon_{78}$       & -0.38782   & -0.39025   & -0.39038 & -0.39044 & -0.39032 & -0.39039 & -0.39039 \\
    $\epsilon_{79}$       & -0.29518   & -0.29498   & -0.29496 & -0.29496 & -0.29514 & -0.29589 & -0.29588 \\
    \bottomrule
    \end{tabular}
\end{table*}

Sixth, 40 X-ray absorption energies for the K-edges and L-edges of the
eight inorganic molecules of ref~\citenum{Kehry.Klopper.ea:Robust.2023}
are studied with the Bethe--Salpeter equation (BSE) in the Green's function
$GW$ approximation using core-valence separation (CVS) for the linear response equations
and the RI approximations throughout the $GW$-BSE workflow.
\cite{Holzer.Klopper:Ionized.2019, Kehry.Klopper.ea:Robust.2023,
Gui.Holzer.ea:Accuracy.2018, Franzke.Holzer.ea:NMR.2022, Holzer.Franzke:Guide.2025}
It was previously shown that CVS is an excellent approximation compared
to the damped response treatment \cite{Kehry.Klopper.ea:Robust.2023,
Kehry.Franzke.ea:Quasirelativistic.2020} and 2c BSE \cite{Kehry.Klopper.ea:Robust.2023}
outperformed 4c DFT. \cite{Kadek.Konecny.ea:X-ray.2015}
The CHYF-PBE functional \cite{Holzer.Franzke:General.2025} (grid size 5a)
served as starting point for the one-shot $G_0W_0$ quasiparticles.
Other computational settings were chosen according to ref~\citenum{Kehry.Klopper.ea:Robust.2023}.
That is, the x2c-TZVPPall-2c orbital basis set \cite{Pollak.Weigend:Segmented.2017}
and a tailored RI auxiliary basis set (see Supporting Dataset) were employed.
The SCF procedure was converged with a threshold of $10^{-9}$\,$E_{\text{h}}$
for the energy and $10^{-8}$ for the root-mean square of the density matrix change.
An ultrafine grid was applied for the exchange-correlation potential. \cite{Holzer:improved.2020}
In detail, we consider the experimental findings for the L-edge of
CrO$_2$Cl$_2$ and VOCl$_3$, \cite{Fronzoni.Stener.ea:X-ray.2009}
and the K -and L-edges of PdCl$_2$, \cite{Sham:X-ray.1983}
SiCl$_4$, \cite{Bozek.Tan.ea:High.1987}
ferrocene, \cite{Ismail.Guillemin.ea:Experimental.2018, Godehusen.Richter.ea:Iron.2017}
TiCl$_4$, TiCl$_3$Cp, and TiCl$_2$Cp$_2$.
\cite{DeBeer-George.Brant.ea:Metal.2005, Wen.Hitchcock:Inner.1993}
Furthermore, we compare SNSO and SAP-X2C to the amfX2C, mmfX2C, and 4c-DKS
results for WCl$_6$. \cite{Konecny.Komorovsky.ea:Exact.2023}
For simplicity, we apply the CVS-TDDFT approach (50 roots) and other
computational settings such as the PBE0-60HF functional and the
uncontracted Dyall-VDZ/aug-cc-pVDZ basis set \cite{Dyall:Relativistic.2007A,
Dyall-Repo, Woon.Dunning:Gaussian.1993} are chosen accordingly.
The CVS ansatz results in a good agreement with the damped response
results of the literature, i.e.\ the deviation for 1e-X2C results herein
and those of the reference are about 1\,eV. We applied very large grids
(grid size 5a) and tight convergence thresholds of $10^{-9}$\,$E_{\text{h}}$
for the energy and $10^{-7}$ for the root-mean square of the density
matrix change. The pseudospectral approximation was utilized for the TDDFT
linear response equations \cite{Holzer:improved.2020} with a convergence
threshold of $10^{-6}$ for the norm of the residuum. 
\cite{Holzer.Klopper:Ionized.2019, Kehry.Franzke.ea:Quasirelativistic.2020}
The molecular structure was taken from ref~\citenum{Konecny.Komorovsky.ea:Exact.2023}.


\section{Results \& Discussion}
\label{sec:results}

\subsection{Total Energies \& Eigenvalues}
\label{subsec:energies}
To begin with, we assess the accuracy of the total SCF energy and spinor eigenvalues of
SNSO and SAP. Results for the gold dimer molecule in Table~\ref{tab:energies} show
that the total energy or the Z-adjusted error $\Delta_{\text{X2C}}$ is not improved with
SNSO compared to 1e-X2C. The deviation to the 4c-DHF reference becomes even larger with
SNSO. This is in strong contrast to SAP-X2C, which substantially reduces the errors.
This confirms the observations for SAP-X2C in ref~\citenum{Surjuse.Valeev:SAP-X2C.2026}.

Spin--orbit splittings and spinor eigenvalues reveal a different picture. For the $s$-type
spinors SNSO only results in a very minor improvement. This is clearly visible for the
$\epsilon_{2s_{1/2}}$ eigenvalues, where 1e-X2C results in a large error of 0.61\,eV
and SNSO reduces this error by only 0.02 to 0.03\,eV. The behavior is due to the
fact that SNSO does not explicitly introduce corrections for $s$-type functions but
the change only arises due to the indirect correction via all other basis functions
and the X2C/SCF diagonalization. In contrast, SAP-X2C leads to a correction of more
than 0.3\,eV. Thus, the deviation to the 4c-DHF reference is halved for this eigenvalue.
For the other eigenvalues listed, SNSO results in a significant improvement over 1e-X2C.
Especially the difference between $\epsilon_{2p_{1/2}}$ and $\epsilon_{2p_{3/2}}$
is substantially reduced from 69.7\,eV to about 67.7\,eV (SNSO), 67.4\,eV (mSNSO),
or 68.0\,eV (SAP), while the difference amounts to 67.6\,eV at the reference level.
Overall, SAP-X2C results in a better agreement of the eigenvalues than SNSO
but both clearly outperform 1e-X2C.

Importantly, total energies are usually of minor importance in practical quantum chemistry
but energy differences are the decisive quantities to characterize chemical reactions.
This allows for error cancellation and the reaction energy for the dissociation of
Au$_2$ into two Au atoms is very similar---despite the large differences in the
total electronic energies. The maximum difference in the total energy is observed
between mSNSO-X2C and SAP-X2C with almost 13\,$E_{\text{h}}$. The reaction energies are,
however, 28.3\,m$E_{\text{h}}$ (1e-X2C), 28.0\,m$E_{\text{h}}$(SNSO-X2C),
28.0\,m$E_{\text{h}}$ (SNSO-DC-X2C), 28.0\,m$E_{\text{h}}$ (mSNSO-X2C), and
28.2\,m$E_{\text{h}}$ (SAP-X2C). At the 4c-DHF level, we obtain 29.0\,m$E_{\text{h}}$.
Therefore, the difference in the reaction energy between the various X2C approaches amounts
to less than 1\,kJ/mol and the deviation from the 4c-DHF reference is at most
2.3\,kJ/mol. Notably, SNSO performs worse than SAP for the energy difference but
1e-X2C performs best due to error cancellation. Considering the thermochemical accuracy
of DFT, the errors are clearly acceptable and all SNSO or SAP-X2C approaches can
be used to assess reaction energies at the DFT level.

\begin{table*}[t]
    \centering
    \caption{Hg M\"ossbauer contact densities and relative contact densities in atomic units ($a_0^{-3}$)
    at the HF level with the uncontracted Dyall-CVQZ+2s1p/aug-cc-pVQZ basis set
    as well as the PBE level with the uncontracted Dyall-CVTZ+2s1p/aug-cc-pVTZ basis set.
    4c-DHF and 4c-DKS results are taken from ref~\citenum{Knecht.Fux.ea:Mossbauer.2011}.
    For the Hg atom, the absolute value is listed, whereas the relative density is given for the molecules.}
    \label{tab:mossbauer}
    \begin{tabular}{@{\extracolsep{8pt}}
    ll
    S[table-format = -7.2]
    S[table-format = -7.2]
    S[table-format = -7.2]
    S[table-format = -7.2]
    S[table-format = -7.2]
    S[table-format = -7.2]@{}
    }
    \toprule
    System & Level & {\text{1e-X2C}} & {\text{SNSO-X2C}} & {\text{SNSO-DC-X2C}} & {\text{mSNSO-X2C}} & {\text{SAP-X2C}} & {\text{4c DIRAC}} \\
    \midrule
    Hg & HF & 2358409.60 & 2357421.08 & 2357265.95 & 2357902.49 & 2359451.67 & 2363827.39 \\
    HgF & HF & -104.68 & -104.93 & -104.95 & -104.92 & -104.84 & -114.54 \\
    HgF$_2$ & HF & -126.55 & -126.94 & -126.97 & -126.91 & -126.86 & -127.85 \\
    HgF$_4$ & HF & -97.26 & -97.56 & -97.60 & -97.41 & -97.37 & -98.09 \\    
    Hg & PBE & 2377165.23 & 2376161.22 & 2376003.65 & 2376650.55 & 2378212.02 & 2372713.57 \\
    HgF & PBE & -71.22 & -71.57 & -71.59 & -71.57 & -71.52 & -75.11 \\
    HgF$_2$ & PBE & -98.54 & -98.96 & -98.99 & -98.98 & -98.88 & -98.74 \\
    HgF$_4$ & PBE & -113.17 & -113.69 & -113.69 & -113.56 & -113.50 & -113.42 \\
    \bottomrule
    \end{tabular}
\end{table*}

\begin{table*}[t]
    \centering
    \caption{NMR isotropic shielding constants in ppm at the Hartree--Fock level with the decontracted Dyall-VDZ (heavy elements)
    and cc-pVDZ (hydrogen) basis. Results for the lighter atoms or molecules and bond distances are listed in the Supporting Dataset.}
    \label{tab:nmr}
    \begin{tabular}{@{\extracolsep{14pt}}
    ll
    S[table-format = 5.1]
    S[table-format = 5.1]
    S[table-format = 5.1]
    S[table-format = 5.1]
    S[table-format = 5.1]
    S[table-format = 5.1]@{}
    }
    \toprule
    System & Nucleus & {\text{1e-X2C}} & {\text{SNSO-X2C}} & {\text{SNSO-DC-X2C}} & {\text{mSNSO-X2C}} & {\text{SAP-X2C}} & {\text{4c-DHF}} \\
    \midrule
    Sr & Sr & 3908.0 & 3907.4 & 3907.3 & 3907.3 & 3907.9 & 3907.5 \\
    Cd & Cd	& 5735.9 & 5735.0 & 5734.9 & 5734.9 & 5735.8 & 5735.2 \\
    Xe & Xe & 7048.4 & 7046.3 & 7046.0 & 7046.0 & 7047.1 & 7046.1 \\
    Ba & Ba & 7517.0 & 7515.1 & 7514.9 & 7514.9 & 7517.0 & 7515.5 \\
    Hg & Hg & 16401.7 & 16394.8 & 16393.7 & 16395.5 & 16401.0 & 16396.9 \\
    Rn & Rn & 20304.0 & 20282.7 & 20279.3 & 20283.4 & 20289.9 & 20281.5 \\
    SnH$_4$ & H & 28.1 & 28.0 & 28.0 & 28.0 & 28.0 & 28.0 \\
	        & Sn & 4202.5 & 4202.8 & 4202.8 & 4202.5 & 4201.9 & 4201.5 \\
    PbH$_4$ & H & 26.9 & 26.8 & 26.9 & 26.8 & 26.8 & 26.8 \\
			& Pb & 13526.1 & 13529.5 & 13528.3 & 13520.6 & 13515.9 & 13510.1 \\
    SbH$_3$ & H & 31.3 & 31.1 & 31.1 & 31.1 & 31.1 & 31.0 \\
			& Sb & 4456.9 & 4456.8 & 4456.9 & 4456.5 & 4454.0 & 4453.7 \\
    BiH$_3$ & H & 26.4 & 26.6 & 26.7 & 26.6 & 26.5 & 26.6 \\
			& Bi & 13611.1 & 13631.7 & 13636.6 & 13626.0 & 13603.5 & 13605.8 \\
    H$_2$Te & H & 39.6 & 39.3 & 39.2 & 39.2 & 39.2 & 39.1 \\
		& Te & 4819.6 & 4814.1 & 4813.2 & 4812.8 & 4811.1 & 4808.9 \\
    H$_2$Po	& H & 46.6 & 46.7 & 46.7 & 46.7 & 46.6 & 46.7 \\
	    & Po & 15830.9 & 15828.8 & 15828.9 & 15821.3 & 15808.6 & 15806.6 \\
    HI & H & 46.6 & 46.2 & 46.1 & 46.1 & 46.1 & 46.0 \\
       & I & 5901.3 & 5893.1  & 5891.7 & 5891.5 & 5892.0 & 5888.6 \\
    HAt & H & 73.8 & 73.4 & 73.3 & 73.3 & 73.2 & 73.2 \\
        & At & 18843.9 & 18800.8 & 18793.8 & 18794.6 & 18798.0 & 18785.0 \\
    \bottomrule
    \end{tabular}
\end{table*}

\subsection{M\"ossbauer Contact Density}
\label{subsec:mossbauer}
Next, we probe the electron density at the nucleus with the M\"ossbauer contact densities
of small mercury compounds as shown in Table~\ref{tab:mossbauer}. For the mercury atom
as the reference system, the absolute contact density is listed, while the relative
contact density is given for the molecules HgF, HgF$_2$, and HgF$_4$. Note that
we show results from the density operator after the picture-change correction in the
main text. The Supporting Dataset further provides the results with the effective
contact density given by the energy derivative with respect to the parameter of the
finite nucleus model. \cite{Filatov:First.2009, Yoshizawa.Filatov.ea:Calculation.2019,
Zhu.Gao.ea:Mossbauer.2020} Trends are essentially the same but the relative effective
contact densities are smaller by about 10\%. Thus, the density operator is considered
for better comparability to the 4c DIRAC reference values. \cite{Knecht.Fux.ea:Mossbauer.2011}

At the HF level, the deviation of the absolute contact density for Hg from the 4c reference
is substantially reduced by SAP-X2C compared to 1e-X2C, while the SNSO approximations
increase the deviation. In contrast, all X2C approaches perform well for relative
contact densities. The impact of the two-electron picture-change error is less
pronounced than for the relative densities. Notably, SAP-X2C is again the top performer
but SNSO also marks an improvement over 1e-X2C. 
For HgF, a comparably large difference is observed between X2C and 4c DIRAC
results. This is due to different formalisms for open-shell systems in DIRAC
\cite{Saue.Bast.ea:DIRAC.2020, Dirac-manual} and TURBOMOLE.
\cite{Balasubramani.Chen.ea:TURBOMOLE.2020, Franzke.Holzer.ea:TURBOMOLE.2023, TURBOMOLE-manual}
The DIRAC SCF module is based on an average-of-configuration formalism similar to
restricted open-shell HF, while TURBOMOLE applies a straightforward Kramers-unrestricted
formalism similar to unrestricted HF, which captures spin polarization in the core region. 
All other systems feature a closed-shell electronic structure and can be treated with a 
Kramers-restricted formalism so that the X2C and 4c Hamiltonians are directly comparable 
without any bias in methodology.

Results at the semilocal DFT level confirm these findings. However, the absolute
contact densities are hardly comparable---most likely due to different grids for the numerical
integration of the DFT exchange-correlation potential. This shows that SNSO and SAP-X2C
perform excellently for relative M\"ossbauer contact densities and are expected to
yield accurate M\"ossbauer isomer shifts, given that electron correlation is treated
at a suitable level. \cite{Zhu.Gao.ea:Mossbauer.2020} SAP-X2C is still to be preferred
due to its better description of the absolute contact density.

\subsection{NMR Shieldings}
\label{subsec:nmr}
NMR spectroscopy is a key property for all-electron approaches due to its great importance
in analytical chemistry. Isotropic NMR shielding constants of atoms and molecules are
listed in Table~\ref{tab:nmr}. Here, the X2C and 4c-DHF calculations both apply a
magnetic balance condition and gauge-including atomic orbitals to allow for a fair
comparison.

For the atoms, the SAP-X2C ansatz results in smaller changes of the isotropic shielding
constants than the SNSO approximations. Especially mSNSO performs excellently and yields
errors of at most 2\,ppm. To compare, the two-electron picture-change error amounts to
at most 22.5\,ppm for the radon atom. The maximum error of SAP-X2C is 8.4\,ppm, marking
a clear improvement over 1e-X2C. The smaller errors of mSNSO compared to SAP-X2C found
for the noble gases are rationalized by the fact that mSNSO was parametrized for the
xenon atom. \cite{Filatov.Zou.ea:Spin-orbit.2013}

For the molecules, all ans\"atze reproduce the pronounced spin--orbit heavy atom on
the light atom effect (SO-HALA) of the tetrel and halogen systems with great agreement.
The reparametrized SNSO approximations and SAP result in the smallest errors, with
SAP performing best overall. At most errors of 0.1\,ppm are observed for these systems.
This marks a truly excellent performance and the errors for the carbon group and pnictogen
compounds are very similar. The shieldings of the heavy elements are also reproduced in
very excellent agreement. Only for the 5p and 6p elements is a pronounced 2ePCE observed.
Here, mSNSO and SAP perform very well and mark the top performers.
The sum of the absolute errors for the heavy elements amounts to 65.6\,ppm and 29.3\,ppm,
respectively. However, 1e-X2C yields a sum of absolute errors of 131.8\,ppm and especially
for the 6p elements substantial deviations are observed. Therefore, SAP-X2C holds the edge and
generally outperforms mSNSO for molecules. Further optimization of SAP basis sets for
X2C calculations might even reduce the errors once more and result in a more systematic
improvement over mSNSO for both atoms and molecules.

Considering the excellent performance of the NMR shielding constants, mSNSO-X2C and
SAP-X2C are expected to accurately treat relativistic effects for NMR spectra,
which are based on relative NMR shielding constants, i.e.\ NMR shifts.
Based on previous work for NMR coupling constants \cite{Franzke:Reducing.2023,
Franzke.Mack.ea:NMR.2021, Yoshizawa:On.2019} we also expect great accuracy for other
NMR properties such as NMR coupling constants and even NMR spin-rotation constants.

\subsection{EPR Hyperfine Coupling \& g-Tensor}
\label{subsec:epr}
\begin{figure}[t]
    \centering
    \includegraphics[width=0.99\linewidth]{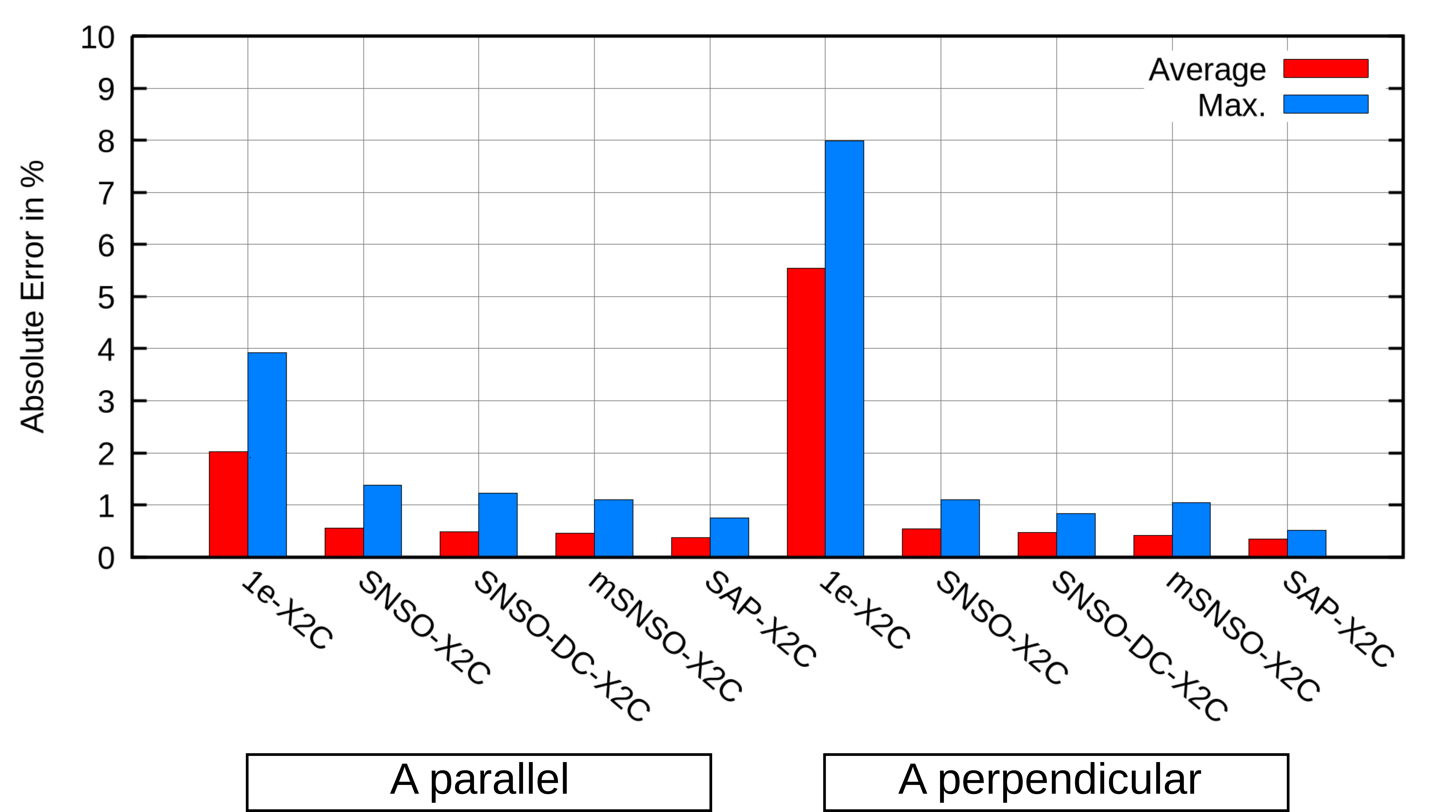}
    \caption{Average and maximum absolute errors for the principal components of the EPR hyperfine
    coupling tensor A of various X2C Hamiltonians with respect to the four-component treatment
    at the PBE0-40HF/Dyall-TZ/IGLO-III level for the set of 17 transition-metal complexes
    compiled in ref~\citenum{Gohr.Hrobarik.ea:Four-Component.2015}. X2C calculations include
    the derivatives of the decoupling and the renormalization matrix as derived in 
    ref~\citenum{Franzke.Yu:Hyperfine.2022}. A-tensor components are labeled according 
    to the principal axis. Complete data are available in the Supporting Dataset.}
    \label{fig:hfc}
\end{figure}

\begin{figure}[ht!]
    \centering
    \includegraphics[width=0.99\linewidth]{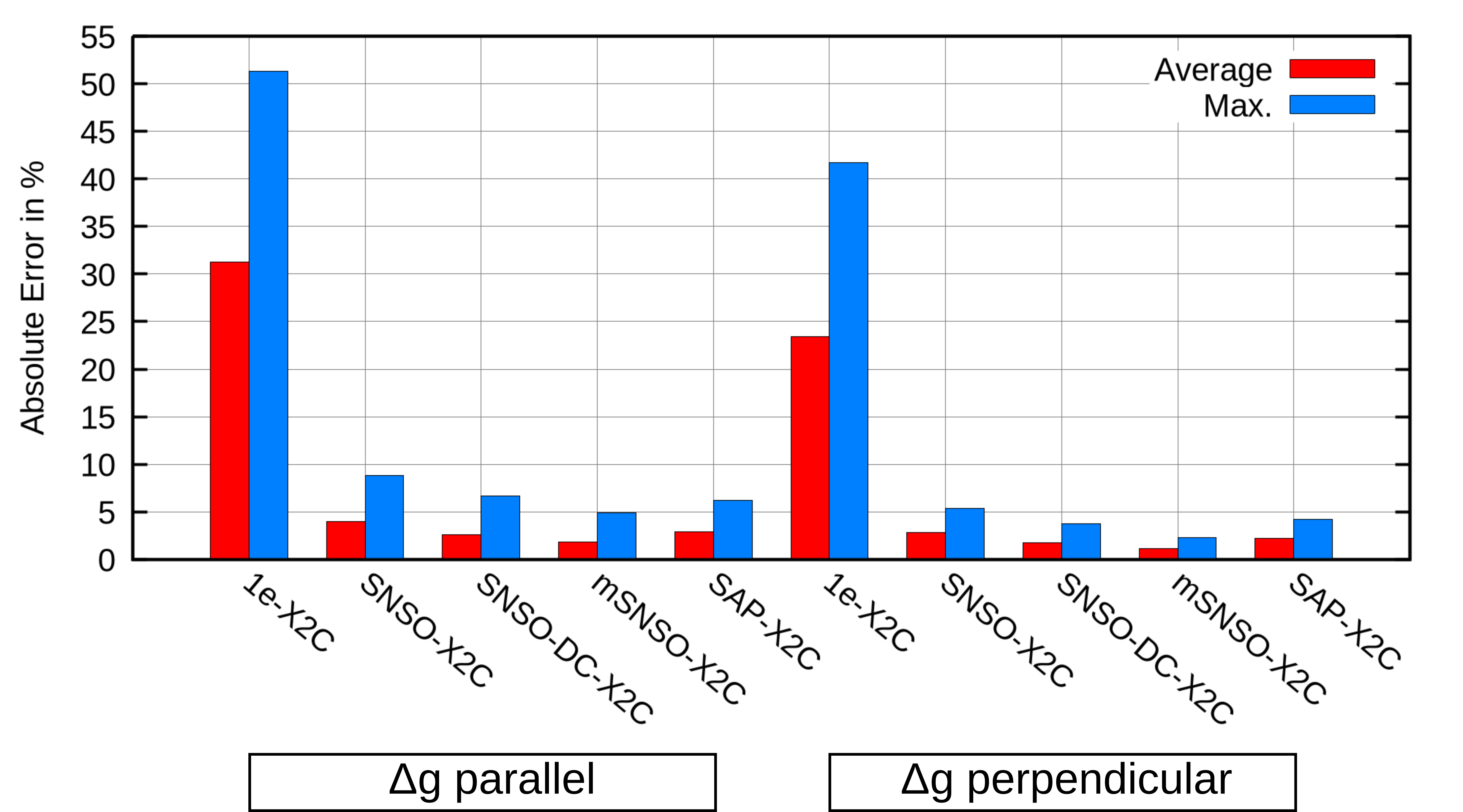}
    \caption{Average and maximum absolute errors for the principal components of the EPR g-shift
    $\Delta g$ in ppt of various X2C Hamiltonians with respect to the four-component treatment
    at the PBE0-40HF/Dyall-TZ/IGLO-III level for the set of 17 transition-metal complexes
    compiled in ref~\citenum{Gohr.Hrobarik.ea:Four-Component.2015}. X2C calculations include
    the derivatives of the decoupling and the renormalization matrix as derived in 
    ref~\citenum{Franzke.Yu:Quasi-Relativistic.2022}. The kinetic balance condition
    is applied with a common gauge origin at the metal center.
    g-shift components are labeled according to the principal axis and g-shifts below
    30\,ppt are neglected in the statistical evaluation.
    Complete data are available in the Supporting Dataset.}
    \label{fig:gtens}
\end{figure}

EPR properties are of great importance in the context of single molecule magnets
and the fundamental characterization of open-shell systems. Here, the 2ePCE is
more pronounced than for closed-shell systems but studies have shown that the
SNSO/mSNSO ansatz accurately considers this error. \cite{Wodynski.Kaupp:Density.2019,
Franzke.Yu:Hyperfine.2022, Franzke.Yu:Quasi-Relativistic.2022} In the present work,
we extend these studies of a test set of 17 transition-metal complexes
\cite{Gohr.Hrobarik.ea:Four-Component.2015} by further considering the SNSO-DC approximation
and the SAP ansatz. Results are displayed in Figs.~\ref{fig:hfc} and \ref{fig:gtens}
for the hyperfine coupling (HFC) constant and the g-shift, respectively.
Due to the molecular point group symmetries, the parallel and perpendicular
principal components are separately assessed. We consider the absolute percent-wise
error (APE) given by
\begin{equation}
    \text{APE}(A) = \frac{|A_{\text{X2C}} - A_{\text{4c}}|}{|A_{\text{4c}}|} \cdot 100
\end{equation}
for the components of the HFC tensor A and likewise for the components of the
g-shift. The percent-wise error is formed, as the absolute values range from
about 100\,MHz to 3050\,MHz for the HFC and from almost 0 to $-$460\,ppt for the g-shift.

For the HFC, the impact of the 2ePCE differs significantly for the parallel and the perpendicular
components. For the parallel component, the mean error amounts to 2.0\% and the maximum error
is 3.9\% found for [ReOBr$_4$]. For the perpendicular component, the corresponding errors read
5.6\% and 8.0\% for [WOBr$_5$]$^{2-}$. Notably, the errors of both components with the SNSO
approximations and the SAP approach are very similar and the mean errors are below 1\%.
Here, SAP-X2C performs best with average errors of 0.4\% and 0.3\%, and maximum errors of
0.7\% and 0.5\% for [ReOBr$_4$] and [WOCl$_4$]$^-$. This is about 0.1\% smaller than for the
SNSO ans\"atze. Among the latter, the mSNSO parameters perform best for the parallel component,
while the SNSO-DC parameters yield smaller errors for the perpendicular component.
Already, the original ones of Boettger are, however, a substantial improvement over the
1e-X2C Hamiltonian.
Considering that the impact of the basis set is much larger for the hyperfine coupling,
\cite{Franzke.Yu:Hyperfine.2022, Cardona.Sole.ea:Exploring.2025} the speed-up of SNSO-X2C
or SAP-X2C compared to the parent four-component approach can be used to employ larger
basis sets with still clearly acceptable errors in the treatment of relativistic effects.

\begin{table}[t]
    \centering
    \caption{Isotropic EPR hyperfine coupling constant $A_{\text{iso}}$ for $^{237}$Np in MHz
    and isotropic g-tensor $g_{\text{iso}}$ of NpF$_6$ at the PBE0-40HF/Dyall-VTZ(Np)/IGLO-III(F)
    level. Results with the amfX2C, eamfX2C, and 4c-DKS Hamiltonians are taken from 
    ref~\citenum{Repisky.Komorovsky.ea:X2C.2025}.
    All other X2C calculations are performed in the present work including the
    derivatives of the decoupling and the renormalization matrix with the kinetic
    balance condition and a common gauge origin at the Np atom. The impact of the
    SAP basis sets is also studied and the SAP basis is given in parentheses.
    Large and small are taken from ref~\citenum{Lehtola.Visscher.ea:Efficient.2020}
    and the compact ones are refitted versions of the large basis set as constructed
    herein.
    Experimental results are taken from ref~\citenum{Hutchison.Weinstock:Paramagnetic.1960}.
    We have added the sign of the A-tensor, as only the absolute value was given in
    the experimental study.}
    \label{tab:npf6}
    \begin{tabular}{@{\extracolsep{16pt}}
    l
    S[table-format = -1.4]
    S[table-format = -4.1]@{}
    }
    \toprule
    & {\text{$g_{\text{iso}}$}} & \text{$A_{\text{iso}}$}  \\
    \midrule
    1e-X2C      & -0.9220 & -1870.5 \\
    SNSO-X2C    & -0.4793 & -1687.2 \\
    SNSO-DC-X2C & -0.4711 & -1683.4 \\
    mSNSO-X2C   & -0.4610 & -1678.3 \\
    SAP-X2C (large)    & -0.4405 & -1672.3 \\
    SAP-X2C (small) & -0.4403 & -1673.2 \\
    SAP-X2C (compact) & -0.4405 & -1672.3 \\
    SAP-X2C (compact2) & -0.4405 & -1672.3 \\
    SAP-X2C (compact3) & -0.4403 & -1672.2 \\
    amfX2C     & -0.4245 & -1660.2 \\
    eamfX2C    & -0.4243 & -1660.1 \\
    \midrule
    4c-DKS      & -0.4248 & -1657.5 \\
    Experiment  & {\text{$-0.604 \pm 0.003$}} & {\text{$-1994 \pm 15 $}} \\
    \bottomrule
    \end{tabular}
\end{table}

Turning to the g-tensor, a drastic increase of the 2ePCE is observed. The mean errors
of 1e-X2C amount to 31.3\% and 23.4\% with maximum errors of 51.3\% and 41.7\%.
The greater impact of the 2ePCE for the g-tensor compared to the HFC may be rationalized
as follows. Generally, the spin--orbit 2ePCE is expected to be more important than the
scalar-relativistic 2ePCE, which is also one of the reasons why no scalar analog of the
SNSO approach was presented in the literature.
The HFC consists of a scalar and a spin--orbit contribution due to the
Fermi-contact, the spin-dipole, and the paramagnetic spin--orbit terms as well as a
purely relativistic term. \cite{Haase.Repisky.ea:Relativistic.2018}
For the HFC of the transition-metal complexes, both the scalar and the spin--orbit
terms are sizable, while the g-tensor is a spin--orbit phenomenon and therefore
we expect the 2ePCE to be larger for the g-tensor than for the HFC of the given
molecular complexes in the test set.

Similar to the HFC, SNSO and SAP greatly reduce the 2ePCE also for the g-tensor.
The mSNSO and SAP correction again perform best. SAP-X2C results in an average
deviation of 2.9\% and 2.2\%, and maximum errors of 6.2\% and 4.2\%.
The mSNSO approximation yields smaller average and maximum errors but it is
a minuscule improvement. That is, both mSNSO and SAP-X2C can be safely used
for an accurate treatment of relativistic effects.

We conclude our assessment for EPR properties with a more extreme example,
namely NpF$_6$, where we also compare the low-cost 2ePCE correction in
SNSO/SAP-X2C to the more complicated amfX2C approach.
As shown in Table~\ref{tab:npf6}, NpF$_6$ can be assigned negative
g-tensor components and HFC components. 1e-X2C performs poorly with an error
of almost 0.5, i.e.\ 500 ppt, in the isotropic g-tensor and an error of more than
200\,MHz for the isotropic HFC constant. SNSO reduces the errors significantly to
about 0.04 and 220\,MHz, respectively. SAP-X2C leads to a further reduction
to 0.02 and 25\,MHz. Due to the pronounced 2ePCE, we consider the EPR properties
of NpF$_6$ as an initial test case for the refitted compact SAP-X2C basis sets, while
a more detailed study becomes relevant for periodic systems in future work.
For NpF$_6$, the different SAP-X2C basis sets perform excellently and EPR results
are almost identical for the compact refitted basis sets. Importantly, all are a
clear and consistent improvement upon both 1e-X2C and all SNSO-X2C variants.
This shows that the diffuse functions may be relevant for the SAP as an SCF initial
guess but a more compact basis can be used when targeting the 2ePCE in X2C.
None of the SAP-X2C approaches yet reaches the almost perfect accuracy of
amfX2C/eamfX2C. Still considering the other computational parameters, a treatment
of the 2ePCE with SAP may be considered sufficiently accurate.
Here, multireference treatments can improve the agreement with the experiment
of $g_{\text{iso}} = -0.604$, \cite{Hutchison.Weinstock:Paramagnetic.1960}
as, for instance, complete active space SCF (CASSCF) using an active space of
96 roots leads to an isotropic g-factor of $g_{\text{iso}} = -0.72$.
\cite{Notter.Bolvin:Optical.2009}
This demonstrates that a consideration of both static electron correlation
and special relativity is important for the EPR properties of NpF$_6$.

\subsection{UV/vis Spectra}
\label{subsec:tddft}
\begin{figure}[ht!]
    \centering
    \includegraphics[width=0.99\linewidth]{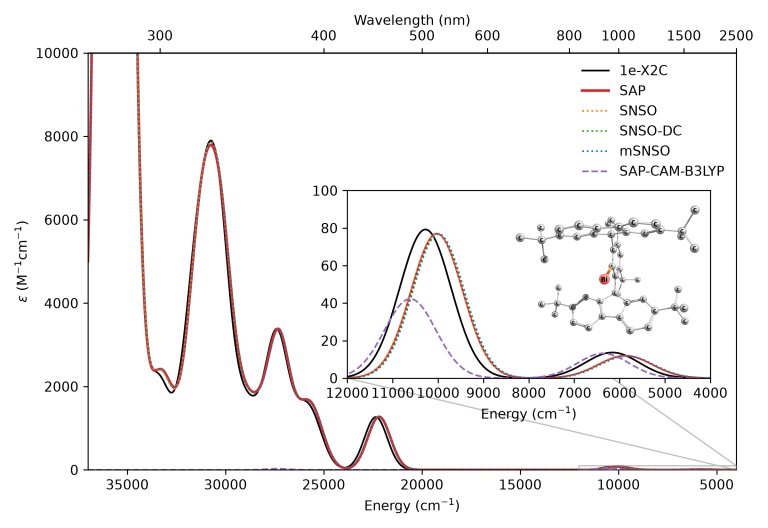}
    \caption{Simulated UV/vis absorption spectrum of the triplet bismuthinidene system of
    ref~\citenum{Pang.Nothling.ea:Synthesis.2023}, computed at the 2c TDDFT (B3LYP) level using different
    two-component relativistic spin--orbit treatments: the bare one-electron X2C Hamiltonian (1e-X2C, black, reference),
    and the approximate local 2ePCE spin--orbit potentials SAP (red), SNSO (orange, dotted), SNSO-DC (green, dotted),
    and mSNSO (blue, dotted). The purple dashed curve (SAP-CAM-B3LYP) additionally illustrates the effect of
    using the range-separated hybrid functional CAM-B3LYP together with the SAP potential correction for the
    near-infrared region. Oscillator strengths (length representation) were convoluted with Gaussian line shapes
    (800 cm$^{-1}$ Gaussian half-width at 1/e height). The inset magnifies the weak near-infrared region
    (12000--4000\,cm$^{-1}$), resolving the two lowest-energy and formally spin-forbidden transitions between
    the spin--orbit-split sublevels of the $S = 1$ triplet ground state (M$_S$ = 0 $\rightarrow$ M$_S$ = $\pm$1).
    See also Table~\ref{tab:zfs} for the zero-field splitting at about 6000\,cm$^{-1}$ with the different
    computational methodologies. The simulated spectrum is in good agreement with the experimental spectrum
    shown in ref~\citenum{Pang.Nothling.ea:Synthesis.2023}.}
    \label{fig:tddft}
\end{figure}

Furthermore, we study the magnetic properties and excitation energies of a molecule
with more than one unpaired spin. We consider the highly popular hydrindacene ligands
which have become a popular choice to prepare low-valent organometallic main group
systems with uncommon optical and magnetic properties.
\cite{Wang.Chen.ea:Synthesis.2025, Wu.Li.ea:triplet.2023, Pang.Nothling.ea:Synthesis.2023,
Al-Said.Spinnato.ea:Direct.2025, Mansikkamaki:Theoretical.2023}
Here, the bismuthinidene system with a ``non-magnetic'' triplet ground state serves as
a prime example. \cite{Pang.Nothling.ea:Synthesis.2023}
For this system, the zero-field splitting is so large that the M$_S = 0$ state is
thermally separated from the M$_S = \pm 1$ states and it behaves like a diamagnetic
molecule at room temperature and cannot be probed with EPR spectroscopy.
However, this giant zero-field splitting can be measured with magneto-optical
infrared spectroscopy \cite{Al-Said.Spinnato.ea:Direct.2025} and can further be calculated
with spin--orbit 2c TDDFT excitation energies due to the approximate axial symmetry. Accordingly,
the excitation energy from the M$_S = 0$ sublevel to the M$_S = \pm 1$ states corresponds to
the axial zero-field splitting parameter $D$, which is listed in Table~\ref{tab:zfs} and
the simulated absorption spectrum is shown in Fig.~\ref{fig:tddft}.

\begin{table}[t]
    \centering
    \caption{First two excitation energies from two-component TDDFT and intensity-weighted average to compute
    the zero-field splitting (ZFS) parameter $D$ in wavenumbers (cm$^{-1}$) of the hydrindacene bismuthinidene
    molecule of ref~\citenum{Pang.Nothling.ea:Synthesis.2023}
    at the two-component DLU-X2C/x2c-TZVPall-2c(Bi)/x2c-SVPall(C,H)/COSMO level with the
    finite nucleus model and the regularized Scalmani--Frisch non-collinear DFT formalism.
    TPSS, TPSSh, and CHYF-B95 apply the current densities from the excited states only,
    full current density treatment is denoted with the prefix ``c'' such as in cTPSS.
    Oscillator strengths in velocity gauge were considered for the weighted average.
    Due to the approximate molecular symmetry, the ZFS is almost fully described by the axial $D$ parameter
    (calculated: $E/D = 0.05$, measured: $ E/D < 0.01$). One-component 1e-X2C complete active space
    SCF (CASSCF) with a spin--orbit mean field (SOMF) approach including a picture-change correction
    for the ZFS are taken from refs~\citenum{Pang.Nothling.ea:Synthesis.2023}
    and \citenum{Al-Said.Spinnato.ea:Direct.2025}. In the CASSCF calculations,
    a small active space of (4,5) and the $n$-electron valence second-order perturbation
    theory (NEVPT2) were applied. Experimental results are taken from ref~\citenum{Al-Said.Spinnato.ea:Direct.2025}.}
    \label{tab:zfs}
    \begin{tabular}{l
    S[table-format = 4.0]
    S[table-format = 4.0]
    S[table-format = 4.0]
    }
    \toprule
    Method & {\text{First Exc.}}  & {\text{Second Exc.}} & {\text{ZFS $D$}} \\
    \midrule
    1e-X2C/B3LYP & 6027 & 6376 & 6173 \\
    SNSO-X2C/B3LYP & 5741 & 6100 & 5892 \\
    SNSO-DC-X2C/B3LYP & 5695 & 6055 & 5846 \\
    mSNSO-X2C/B3LYP & 5707 & 6067 & 5858 \\
    SAP-X2C/B3LYP & 5737 & 6095 & 5889 \\
    SAP-X2C/CAM-B3LYP & 6270 & 6408 & 6342 \\ 
    SAP-X2C/TPSS & 5656 & 5874 & 5716 \\
    SAP-X2C/cTPSS & 6094 & 6289 & 6154 \\
    SAP-X2C/TPSSh & 5791 & 6005 & 5853 \\
    SAP-X2C/cTPSSh & 6214 & 6408 & 6272 \\
    SAP-X2C/CHYF-B95 & 4161 & 4300 & 4230 \\
    SAP-X2C/cCHYF-B95 & 4166 & 4321 & 4221 \\
    \midrule
    X2C/CASSCF-NEVPT2 & & & 4523 \\
    Experiment & & & 5422 \\
    \bottomrule
    \end{tabular}
\end{table}

For the magnitude of the zero-field splitting, all spin--orbit X2C-DFT methods perform well
and lead to an excellent agreement with the experiment. We obtain a zero-field splitting of about 6000\,cm$^{-1}$
with all X2C Hamiltonians compared to the experiment of 5422\,cm$^{-1}$. Treatment of the 2ePCE
with SNSO or SAP leads to a smaller zero-field splitting by 300\,cm$^{-1}$ as the excitation
energies are generally slightly red-shifted, as evident from the simulated absorption
spectrum---especially at low absorption energies.
Results with SNSO/SAP only differ by about 50\,cm$^{-1}$ and therefore
the particular choice of the 2ePCE correction is far less important than just accounting for
the 2ePCE. Here, the SNSO and SAP-X2C results range from 5846 to 5892\,cm$^{-1}$ compared to
a zero-field splitting of 6173\,cm$^{-1}$ with 1e-X2C.

Notably, the two-component spin--orbit B3LYP methods lead to a better agreement with the
experiment than the one-component X2C-based CASSCF-NEVPT2 approach with a SOMF correction.
\cite{Al-Said.Spinnato.ea:Direct.2025, Pang.Nothling.ea:Synthesis.2023}
CHYF-B95 and TPSS further improve the agreement with the experiment. For TPSS, the
excellent results can, however, be attributed to error-cancellation, as the full
current density treatment substantially worsens the agreement with the experiment.
As observed previously, \cite{Holzer.Franke.ea:Current.2022} including the current
density for the excited states is required for stability. Fully neglecting the
currents leads to imaginary eigenvalues and the response equations do not converge.
Application of the range-separated functional CAM-B3LYP increases
zero-field splitting by about 450\,cm$^{-1}$.
This significantly worsens the agreement with the experiment.
So, a better agreement with the experiment likely requires an improved spatial admixture of
exact exchange with local hybrids, as range-separated hybrids are not flexible enough.
We note in passing that for the truncated phenylbismuth molecule, multireference
calculations with a large active space of (10,10) lead to $D = 3122$\,cm$^{-1}$ (CASSCF)
and $D = 3391$\,cm$^{-1}$ (CASSCF-NEVPT2) as discussed in 
ref~\citenum{Mansikkamaki:Theoretical.2023}.

For the complete spectrum, no qualitative changes are observed between 1e-X2C and the
2ePCE-corrected X2C variants. The main change is the slight red shift for absorption
energies between 5000 and 25000\,cm$^{-1}$ and the experimental spectra are always
well reproduced with spin--orbit X2C-DFT.
In the experiment, \cite{Pang.Nothling.ea:Synthesis.2023} absorption energies with
weak intensities are found at approximately 10000, 20500, and 25500\,cm$^{-1}$,
whereas intense Gaussian bands are fitted for about 28000 and 30000\,cm$^{-1}$.
The sixth Gaussian band with the largest intensity is at around 32000\,cm$^{-1}$.
In the DFT-based spectrum, the intensity of the absorption band at around
20000\,cm$^{-1}$ is overestimated and the bands are generally blue-shifted
compared to the experiment. Notably, the absorption spectra simulated with
SAP-X2C and the different SNSO-X2C approximations are virtually indistinguishable.

\subsection{Core-Level Spectra: K-Edges \& L-Edges}
\label{subsec:xas}
Core-level spectra often show larger relativistic effects and consequently the
2ePCE is expected to be more pronounced than for UV/vis spectra, which are associated
with the excitations in the valence region. Therefore, we apply SAP-X2C to X-ray
absorption spectroscopy (XAS), where 1e-X2C is known to result in substantial errors
for relativistic effects. \cite{Konecny.Komorovsky.ea:Exact.2023}

As outlined by Table~\ref{tab:xas_statistics}, standard 1e-X2C is uniformly the worst
performer (highest MSD, MAD, and STD) for predicting core excitations in X-ray absorption
spectroscopy, confirming the corrections are needed. SNSO-DC and mSNSO are the 
best performers in both blocks, cutting MAD roughly in half relative to 1e-X2C and giving the 
tightest STD with 0.8\,eV at the K-edges and 1.0--1.1\,eV at the L-edges.
The original SNSO sits between these and the uncorrected 1e-X2C. In the case of $G_0W_0$-BSE
core excitations, SAP is the weakest of the corrected methods---its K-edge MAD (1.351\,eV)
approaches that of 1e-X2C, and at the L-edge it matches the MAD of 1e-X2C (0.65\,eV) while
having the highest MAPD (8.75\%) of any method. This is, however, caused by a single outlier
in the L$_{2,3}$-edge of PdCl$_2$, which has been shown to be notoriously hard to describe correctly.
\cite{Kehry.Klopper.ea:Robust.2023} Thus, the comparison to the experimental results is
of great importance in practice but may lead to a bias for the assessment of SAP vs.\ SNSO.

\begin{table*}[t]
\centering
\caption{Statistical summary of the deviations of 1e-X2C, SNSO-X2C, SNSO-DC-X2C, mSNSO-X2C, and SAP-X2C
from the experimental reference values resolved by the edges in X-ray absorption spectroscopy. Calculations
were performed at the $GW$-BSE@CHYF-PBE level of theory using the x2c-TZVPPall-2c basis set for all atoms
and the core-valence separation approximation. 10 data points are available for the K+L$_1$-edge, whereas
32 are available for the L$_{2,3}$-edge. MSD denotes the mean signed deviation, MAD the mean absolute
deviation, MAPD the mean absolute percent-wise deviation, and STD the standard deviation. Max.\ Abs.\ Err.\ refers
to the maximum absolute error.}
\label{tab:xas_statistics}
\begin{tabular}{@{\extracolsep{10pt}}
l
l
S[table-format = 2.2]
S[table-format = 2.2]
S[table-format = 2.2]
S[table-format = 2.2]
S[table-format = 2.2]@{}
}
\toprule
Edge & Statistic & {\text{1e-X2C}} & {\text{SNSO-X2C}} & {\text{SNSO-DC-X2C}} & {\text{mSNSO-X2C}} & {\text{SAP-X2C}} \\
\midrule
\multirow{6}{*}{{\text{1s$+$2s (K+L$_1$)}}}
 & MSD (eV)             & 1.81 & 0.66 & 0.57 & 0.57 & 0.95 \\
 & MAD (eV)             & 1.81 & 1.02 & 0.83 & 0.77 & 1.35 \\
 & MAPD (\%)            & 5.16 & 3.83 & 5.84 & 5.45 & 4.26 \\
 & STD (eV)             & 1.36 & 1.19 & 0.84 & 0.78 & 1.52 \\
 & Max.\ Abs.\ Err.\ (eV) & 4.43 & 2.83 & 2.18 & 2.20 & 3.87 \\
\midrule
\multirow{6}{*}{{\text{2p$_{1/2}+$2p$_{3/2}$ (L$_{2}+$L$_3$})}}
 & MSD (eV)             & 0.21 & 0.06 & 0.01 & 0.10 & 0.06 \\
 & MAD (eV)             & 0.65 & 0.50 & 0.46 & 0.50 & 0.65 \\
 & MAPD (\%)            & 8.68 & 7.29 & 7.02 & 7.15 & 8.75 \\
 & STD (eV)             & 1.94 & 1.13 & 1.00 & 1.12 & 1.49 \\
 & Max.\ Abs.\ Err.\ (eV) & 10.48 & 5.43 & 4.55 & 5.07 & 7.20 \\
\bottomrule
\end{tabular}
\end{table*}

Absolute deviations are roughly twice as large for the K-edge (1s$+$2s) block as for the
L-edge (2p) block across every method, e.g.\ MAD 1.81\,eV vs. 0.65\,eV for 1e-X2C, 0.83\,eV
vs.\ 0.46\,eV for SNSO-DC. This is expected given the much larger absolute binding energies
at the K-edge. Once normalized, the picture flips: MAPE is consistently higher for the
2p block (7-9\%) than for 1s$+$2s (4-6\%), since the same absolute error represents a larger
fraction of the smaller L-edge energies. The 2p block also contains the single worst
outlier overall---1e-X2C's maximum error there (10.48\,eV) is more than double its worst
K-edge error (4.43\,eV), pointing to a few problematic L-edge cases that the uncorrected
method handles especially poorly. Parts of the deviation can also be attributed 
to the $G_0W_0$ reference, but a systematic improvement of applying relativistic two-electron
contributions can be seen. The results in Table~\ref{tab:xas_statistics} therefore suggest that
either SNSO or SAP corrections should be used, with both yielding corrections that statistically 
steer the results towards the correct direction.

\begin{table}[t]
    \centering
    \caption{Metal XAS L$_{2,3}$-edge positions and corresponding spin--orbit splittings $\Delta$SO in eV
    for WCl$_6$ with different relativistic Hamiltonians at the PBE0-60HF/Dyall-VDZ(W)/aug-cc-pVDZ(Cl)
    level with damped response (DR) TDDFT or core-valence separation (CVS) linear-response TDDFT.
    Basis sets were employed in fully decontracted form. DR-TDDFT results are taken from
    ref~\citenum{Konecny.Komorovsky.ea:Exact.2023} and experimental data are taken from
    ref~\citenum{Jayarathne.Chandrasekaran.ea:X-ray.2014}. Here, (e)amfX2C means that
    amfX2C and eamfX2C lead to the same XAS results at the given precision.}
    \label{tab:wcl6}
    \begin{tabular}{@{\extracolsep{4pt}}
    lc
    S[table-format = 5.1]
    S[table-format = 5.1]
    S[table-format = 4.1]@{}
    }
    \toprule
    Approach & TDDFT & {\text{L$_3$}} & {\text{L$_2$}} & {\text{$\Delta$SO}} \\
    \midrule
    1e-X2C & DR       & 10185.9 & 11580.6 & 1394.7 \\
    1e-X2C & CVS      & 10186.9 & 11581.6 & 1394.7 \\
    SNSO-X2C & CVS    & 10198.0 & 11552.0 & 1354.0 \\
    SNSO-DC-X2C & CVS & 10199.8 & 11547.4 & 1347.6 \\
    mSNSO-X2C & CVS   & 10199.8 & 11547.5 & 1347.7 \\
    SAP-X2C & CVS     & 10203.9 & 11562.6 & 1358.7 \\
    (e)amfX2C & DR    & 10207.3 & 11561.6 & 1354.3 \\
    mmfX2C & DR        & 10207.3 & 11561.7 & 1354.4 \\
    \midrule
    4c-DKS & DR      & 10207.3 & 11561.7 & 1354.4 \\
    Experiment &     & 10212.2 & 11547.0 & 1334.8 \\
    \bottomrule
    \end{tabular}
\end{table}

The comparison to the experimental findings may be biased by the
error of the complete computational methodology. Therefore, we
also study the performance of SNSO and SAP-X2C for molecules with
available four-component results as reference. 
Here, we study the WCl$_6$ molecule due to its very pronounced 2ePCE.
\cite{Konecny.Komorovsky.ea:Exact.2023}
Results for the metal L$_{2,3}$-edge positions and spin--orbit splittings
are shown in Table~\ref{tab:wcl6}. This reveals a different picture
than the previous comparison to experimental results. SAP-X2C performs
best for the edge positions---especially for the L$_2$-edge.
The SNSO approximations are a clear improvement over the bare 1e-X2C
Hamiltonian but the individual edge positions show a larger deviation
from the 4c reference. Spin--orbit splittings are greatly improved
by all 2ePCE corrections and the original SNSO ansatz shows a minuscule
error of only 0.4\,eV towards the 4c-DKS result. Considering the
individual edge positions, this is, however, due to error cancellation,
as both edge positions differ by about 9\,eV from those of 4c-DKS.

Taken together, SAP-X2C corrects the 2ePCE in the XAS metal edge positions 
more accurately than any of the SNSO-X2C versions, and is thus superior 
in the treatment of special relativity.

\section{Summary \& Conclusion}
\label{sec:conclusion}
We have generalized and thoroughly assessed the SAP-X2C approach
to treat the two-electron picture-change error in X2C for molecular
spectroscopy. Especially for EPR properties, the two-electron
picture-change error is significant and has to be accounted for.
The accuracy of SAP-X2C is compared to the SNSO-X2C approximations,
which come with similar computational costs.
Overall, the mSNSO approximation by the Cremer group and SAP-X2C
perform rather similarly with SAP-X2C yielding slightly lower errors
compared to the four-component methodology.
Thanks to its very simple implementation based on integrals for
the finite nucleus model, its high efficiency, and the publicly 
available basis sets, the SAP-X2C ansatz should be made the default
X2C Hamiltonian for molecular properties and computational spectroscopy.

\appendix
\section{Appendix: Local-Hybrid Energy and Its Second Variation}
\allowdisplaybreaks
We summarize the mathematical formulation of the two-component local-hybrid
kernel used for Kramers-restricted and general non-collinear reference states.
The two formulations share the same local-hybrid product rule and differ only
in the transformation from physical charge and spin densities to the
spin-resolved variables of a collinear density functional.  Special attention
is paid to the response of the exact-exchange energy density and to the
gauge-invariant kinetic-energy density required by current-dependent
meta-generalized gradient approximation (meta-GGA) functionals.
Global and range-separated hybrid functionals of the GGA and meta-GGA rung
follow as the straightforward simplification.

At each point in space, the local-hybrid energy density is written as
\begin{equation}
  \varepsilon_{\mathrm{LH}}
  =(1-a)e_{\mathrm{x}}^{\mathrm{XC}}
   +a\,e_{\mathrm{x}}^{\mathrm{exact}}
   +e_{\mathrm{c}}^{\mathrm{XC}}
  \label{eq:lh}
\end{equation}
Here $e_{\mathrm{x}}^{\mathrm{XC}}$ and
$e_{\mathrm{c}}^{\mathrm{XC}}$ denote the semilocal exchange and correlation
energy densities, $a$ is a local mixing function, and
$e_{\mathrm{x}}^{\mathrm{exact}}$ is the exact-exchange energy density.  The
semilocal quantities depend on the seven spin-resolved variables
\begin{equation}
  \ve u=
  \left(
  \rho_\alpha,\rho_\beta,
  \gamma_{\alpha\alpha},\gamma_{\alpha\beta},\gamma_{\beta\beta},
  \tau_\alpha,\tau_\beta
  \right),
  \qquad
  \gamma_{\sigma\sigma'}
  =\nabla\rho_\sigma\cdot\nabla\rho_{\sigma'}
  \label{eq:u}
\end{equation}

It is useful first to hold $e_{\mathrm{x}}^{\mathrm{exact}}$ fixed.  With
subscripts denoting derivatives with respect to components of $\ve u$, one
obtains
\begin{align}
  G_i
  &=(1-a)e_{\mathrm{x},i}^{\mathrm{XC}}
    +\left(e_{\mathrm{x}}^{\mathrm{exact}}
           -e_{\mathrm{x}}^{\mathrm{XC}}\right)a_i
    +e_{\mathrm{c},i}^{\mathrm{XC}}
  \label{eq:g1}\\
  G_{ij}
  &=(1-a)e_{\mathrm{x},ij}^{\mathrm{XC}}
    +\left(e_{\mathrm{x}}^{\mathrm{exact}}
           -e_{\mathrm{x}}^{\mathrm{XC}}\right)a_{ij}
    +e_{\mathrm{c},ij}^{\mathrm{XC}}
    -a_i e_{\mathrm{x},j}^{\mathrm{XC}}
    -a_j e_{\mathrm{x},i}^{\mathrm{XC}}
  \label{eq:g2}
\end{align}
Equation~\eqref{eq:g2} is the common local-hybrid Hessian in both
implementations.

Let $\ve p$ denote the physical density variables and let $\delta$ indicate a
response direction.  The local potential associated with a primitive $p_\mu$
is
\begin{equation}
  V_\mu^{\mathrm{loc}}=G_i\,u_{i,\mu}
\end{equation}
Its variation is
\begin{equation}
  \delta V_\mu^{\mathrm{loc}}
  =G_{ij}\,u_{i,\mu}\,\delta u_j
   +G_i\,\delta u_{i,\mu}
   +(\delta e_{\mathrm{x}}^{\mathrm{exact}})\,a_i u_{i,\mu}
  \label{eq:local-kernel}
\end{equation}
with
\begin{equation}
  u_{i,\mu}=\frac{\partial u_i}{\partial p_\mu},
  \qquad
  \delta u_{i,\mu}
  =\frac{\partial^2u_i}{\partial p_\mu\partial p_\nu}
    \,\delta p_\nu
\end{equation}
The first term in eq~\eqref{eq:local-kernel} is the ordinary Hessian
contraction, the second is the curvature of the transformation
$\ve p\mapsto\ve u$, and the third accounts for the response of the
exact-exchange energy density.

The corresponding nonlocal exact-exchange operator satisfies the elementary
product rule
\begin{equation}
  \delta\!\left(a\hat V_x\right)
  =a\,\delta\hat V_x+(\delta a)\hat V_x,
  \qquad
  \delta a=a_i\,\delta u_i
  \label{eq:nonlocal-kernel}
\end{equation}
Equations~\eqref{eq:local-kernel} and \eqref{eq:nonlocal-kernel} together form
the complete local-hybrid kernel.

For a self-adjoint linear exchange map $\mathcal A$, the exact-exchange
energy density may be represented schematically as
\begin{equation}
  e_{\mathrm{x}}^{\mathrm{exact}}
  =\frac12\langle D,\mathcal A[D]\rangle
\end{equation}
Consequently,
\begin{align}
  \delta e_{\mathrm{x}}^{\mathrm{exact}}
  &=\frac12\left(
    \langle\delta D,\mathcal A[D]\rangle
    +\langle D,\mathcal A[\delta D]\rangle\right)
  \label{eq:exx1}\\
  \delta_1\delta_2 e_{\mathrm{x}}^{\mathrm{exact}}
  &=\frac12\left(
    \langle\delta_1D,\mathcal A[\delta_2D]\rangle
    +\langle\delta_2D,\mathcal A[\delta_1D]\rangle\right)
  \label{eq:exx2}
\end{align}
The symmetrized form makes the Hermiticity of the response kernel explicit.

\subsection{Two-Component Charge and Spin Variables}

The local density and kinetic-energy-density matrices are expanded in Pauli
form,
\begin{equation}
  R=\frac12\left(n\,I+\ve m\cdot\ve\sigma\right),
  \qquad
  Q=\frac12\left(T\,I+\ve K\cdot\ve\sigma\right)
  \label{eq:pauli}
\end{equation}
Here $R=R(\ve r)$ is the Hermitian $2\times2$ spin-density matrix and
$Q=Q(\ve r)$ is the corresponding kinetic-energy-density matrix at the
spatial point $\ve r$.  The symbol $I$ denotes the $2\times2$ identity
\begin{equation}
  I=\begin{pmatrix}1&0\\0&1\end{pmatrix}
\end{equation}
and
$\ve\sigma=(\sigma_x,\sigma_y,\sigma_z)$ is the vector of Pauli spin matrices,
\begin{equation}
  \sigma_x=\begin{pmatrix}0&1\\1&0\end{pmatrix},
  \sigma_y=\begin{pmatrix}0&-i\\i&0\end{pmatrix},
  \sigma_z=\begin{pmatrix}1&0\\0&-1\end{pmatrix}
\end{equation}
The dot denotes contraction over the three spin components, for example
$\ve m\cdot\ve\sigma=\sum_{k=x,y,z}m_k\sigma_k$.  The scalar fields and
spin-vector fields in eq~\eqref{eq:pauli} are equivalently defined by
\begin{align}
  n&=\Tr R,             &m_k&=\Tr(R\sigma_k) \\
  T&=\Tr Q,             &K_k&=\Tr(Q\sigma_k)
\end{align}
Thus $n$ is the total particle density, $\ve m=(m_x,m_y,m_z)$ is the
magnetization density, $T$ is the total kinetic-energy density, and
$\ve K=(K_x,K_y,K_z)$ is its spin-vector counterpart.  The two matrices may
also be written explicitly as
\begin{align}
  R= & \frac12
  \begin{pmatrix}
    n+m_z&m_x-i m_y\\
    m_x+i m_y&n-m_z
  \end{pmatrix} \\
  Q= & \frac12
  \begin{pmatrix}
    T+K_z&K_x-iK_y\\
    K_x+iK_y&T-K_z
  \end{pmatrix}
\end{align}

For each spatial direction $\lambda\in\{x,y,z\}$, the paramagnetic-current
matrix is decomposed in the same way,
\begin{equation}
  J_\lambda
  =\frac12\left(j_{0\lambda}I
        +\sum_{k=x,y,z}j_{v\lambda k}\sigma_k\right)
\end{equation}
Here $j_{0\lambda}=\Tr J_\lambda$ is the $\lambda$ component of the charge
current, while $j_{v\lambda k}=\Tr(J_\lambda\sigma_k)$ is the current flowing
in spatial direction $\lambda$ with spin component $k$.

Using these fields, the physical primitive set is
\begin{equation}
  \ve p=
  \left(n,\nabla n,\ve m,\nabla\ve m,T,\ve K,
        \ve j_0,\ve j_v\right)
  \label{eq:p}
\end{equation}
where $\nabla n=(\partial_\lambda n)$ is the spatial gradient of the charge
density and $\nabla\ve m=(\partial_\lambda m_k)$ is a rank-two tensor
containing the spatial gradient of every magnetization component.  Likewise,
$\ve j_0=(j_{0\lambda})$ contains three charge-current components and
$\ve j_v=(j_{v\lambda k})$ denotes the nine-component spin-current tensor.
Repeated $k$ and $\lambda$ indices are summed unless stated otherwise.
Altogether, eq~\eqref{eq:p} contains 32 real ground-state fields; their
linear-response variations may in general be complex.

\subsection{Kramers-Restricted Reference}

For a Kramers-restricted reference, $\ve m=\ve K=0$.  Define the per-spin
quantities
\begin{equation}
  t=\frac n2,\qquad \mathcal T=\frac T2
\end{equation}
and the rotational invariants
\begin{align}
  \ve q&=(t,z,\mathcal T),
  &z&=|\nabla t|^2,
  \label{eq:q}\\
  \ve b_k&=(m_k,y_k,K_k),
  &y_k&=\nabla t\cdot\nabla m_k,
  \label{eq:b}\\
  x&=\sum_k|\nabla m_k|^2 
  \label{eq:x}
\end{align}
Choosing an arbitrary local spin axis gives the collinear transformation
\begin{align}
  \rho_\alpha&=t+m,
  &\rho_\beta&=t-m,
  \nonumber\\
  \gamma_{\alpha\alpha}&=z+2y+x,
  &\gamma_{\alpha\beta}&=z-x,
  &\gamma_{\beta\beta}&=z-2y+x,
  \label{eq:closed-map}\\
  \tau_\alpha&=\mathcal T+K,
  &\tau_\beta&=\mathcal T-K .
  \nonumber
\end{align}
Spin-rotation invariance promotes the collinear triplet
$(m,y,K)$ to the three vectors $\ve b_k$.  At the unpolarized reference,
the Hessian therefore separates into a scalar block and three identical
vector blocks,
\begin{equation}
  \delta^2\varepsilon_{\mathrm{LH}}
  =\delta\ve q^{\,T}B\,\delta\ve q
   +\sum_{k=x,y,z}\delta\ve b_k^{\,T}S\,\delta\ve b_k
   +G_z\,\delta^2z+G_x\,\delta^2x .
  \label{eq:closed-hessian}
\end{equation}
The curvature terms follow from
\begin{align}
  \delta z&=2\nabla t\cdot\delta\nabla t,
  &\delta^2z&=2|\delta\nabla t|^2,
  \nonumber\\
  \delta y_k&=\nabla t\cdot\delta\nabla m_k,
  &\delta^2y_k&=2\,\delta\nabla t\cdot\delta\nabla m_k,
  \label{eq:closed-curvature}\\
  \delta x&=0,
  &\delta^2x&=2\sum_k|\delta\nabla m_k|^2
  \nonumber
\end{align}
The term proportional to $\delta^2y_k$ vanishes because the first derivative
with respect to a spin vector is zero at a spin-isotropic reference.

\subsection{Current-Dependent Kinetic Energy}

Local hybrid functionals introduce the kinetic energy density $\tau$, 
similar to a current-dependent meta-GGA.
The kinetic energy density therefore needs to be replaced by the gauge-invariant expression
\cite{Dobson:Alternative.1993,Becke:Current.2002,Pausch.Holzer:Linear.2022}
\begin{equation}
  \overline{\mathcal T}
  =\mathcal T-\frac{|\ve j|^2}{2t}.
  \label{eq:kr-current}
\end{equation}
Its first variation is
\begin{equation}
  \delta\overline{\mathcal T}
  =\delta\mathcal T
   -\frac{\ve j\cdot\delta\ve j}{t}
   +\frac{|\ve j|^2}{2t^2}\,\delta t
  \label{eq:kr-current-var}
\end{equation}
Substitution of eq~\eqref{eq:kr-current-var} into
eq~\eqref{eq:local-kernel} generates both the density--current coupling and
the current--current part of the kernel.  Because both arise from the same
scalar quantity \eqref{eq:kr-current}, the resulting mixed derivatives are
mutual adjoints. This results can be achieved by closely following 
ref~\citenum{Pausch.Holzer:Linear.2022}, and has been proven to be invariant
in magnetic fields for second derivatives.
We also refer to refs~\citenum{Holzer.Franke.ea:Current.2022} and
\citenum{Franzke.Pausch.ea:Application.2025} for a study on current densities
with spin--orbit coupling in a relativistic framework.

\subsection{Scalmani--Frisch Transformation}

As the second derivatives are very sensitive to numerical noise already, a stable 
scheme needs to be implemented. We choose the formalism of Scalmani and Frisch, 
\cite{Scalmani.Frisch:New.2012} as extensive testing has shown no 
issues with the provided auxiliary variables.
For a finite magnetization $m=|\ve m|$, ref~\citenum{Scalmani.Frisch:New.2012} introduces
\begin{align}
  z&=|\nabla n|^2,
  &q&=\sum_k|\nabla m_k|^2,
  \nonumber\\
  c_k&=\nabla n\cdot\nabla m_k,
  &\eta&=|\ve c|,
  \label{eq:nc-invariants}\\
  s_\eta&=\operatorname{sign}(\ve m\cdot\ve c),
  &s_\tau&=\operatorname{sign}(\ve m\cdot\ve K)
  \nonumber
\end{align}
The spin-resolved variables supplied to the collinear functional are
\begin{align}
  \rho_\alpha&=\frac{n+m}{2},
  &\rho_\beta&=\frac{n-m}{2},
  \label{eq:nc-rho}\\
  \gamma_{\alpha\alpha}
    &=\frac{z+q}{4}+\frac{s_\eta\eta}{2},
  &\gamma_{\alpha\beta}
    &=\frac{z-q}{4},
  &\gamma_{\beta\beta}
    &=\frac{z+q}{4}-\frac{s_\eta\eta}{2},
  \label{eq:nc-gamma}\\
  \tau_\alpha^{(0)}
    &=\frac{T+s_\tau|\ve K|}{2},
  &\tau_\beta^{(0)}
    &=\frac{T-s_\tau|\ve K|}{2}
  \label{eq:nc-tau}
\end{align}
The signs are fixed at the reference state when directional derivatives are
taken.

Let $\ve e=\ve m/m$.  The elementary variations required by the chain rule
are then
\begin{align}
  \delta m&=\ve e\cdot\delta\ve m,
  &\delta\ve e
    &=\frac{\delta\ve m-\ve e\,\delta m}{m},
  \nonumber\\
  \delta z&=2\nabla n\cdot\delta\nabla n,
  &\delta q&=2\sum_k\nabla m_k\cdot\delta\nabla m_k,
  \nonumber\\
  \delta c_k
    &=\delta\nabla n\cdot\nabla m_k
      +\nabla n\cdot\delta\nabla m_k,
  &\delta\eta&=\widehat{\ve c}\cdot\delta\ve c,
  \label{eq:nc-variations}\\
  \delta|\ve K|&=\widehat{\ve K}\cdot\delta\ve K
  \nonumber
\end{align}
Together with eqs~\eqref{eq:nc-rho}--\eqref{eq:nc-tau}, these relations
provide $\delta u_i$ and $\delta u_{i,\mu}$ in
eq~\eqref{eq:local-kernel}.  Equivalently, the composite Hessian is
\begin{equation}
  \frac{\partial^2 G}{\partial p_\mu\partial p_\nu}
  =G_{ij}
    \frac{\partial u_i}{\partial p_\mu}
    \frac{\partial u_j}{\partial p_\nu}
   +G_i
    \frac{\partial^2u_i}{\partial p_\mu\partial p_\nu}
  \label{eq:composite}
\end{equation}
The second term contains the response of the local spin frame and is
essential for transverse spin perturbations.

The generalization to the regularized Scalmani--Frisch auxiliary variables
\cite{Komorovsky.Cherry.ea:Four-component.2019} is straightforward and
also implemented. The same holds for the modified variables of
ref~\citenum{Egidi.Sun.ea:Two-Component.2017}.

\subsection{Gauge-Invariant Eigenvalue Currents}

For each spatial direction $\lambda$, define
\begin{equation}
  j_{\alpha\lambda}
  =j_{0\lambda}+s_{j\lambda}|\ve j_{v\lambda}|,
  \qquad
  j_{\beta\lambda}
  =j_{0\lambda}-s_{j\lambda}|\ve j_{v\lambda}|
  \label{eq:eigen-currents}
\end{equation}
where
\begin{equation}
  s_{j\lambda}
  =\operatorname{sign}(\ve m\cdot\ve j_{v\lambda})
\end{equation}
The gauge-corrected kinetic eigenvalues are
\begin{align}
  \overline\tau_\alpha
  &=\tau_\alpha^{(0)}
    -\frac{1}{2\rho_\alpha}\sum_\lambda j_{\alpha\lambda}^2
  \label{eq:tau-alpha}\\
  \overline\tau_\beta
  &=\tau_\beta^{(0)}
    -\frac{1}{2\rho_\beta}\sum_\lambda j_{\beta\lambda}^2
  \label{eq:tau-beta}
\end{align}
For differentiation near a vanishing spin-current norm, the squares are kept
in the equivalent expanded form
\begin{align}
  S_\alpha
  &=\sum_\lambda\left[
    j_{0\lambda}^2+|\ve j_{v\lambda}|^2
    +2s_{j\lambda}j_{0\lambda}|\ve j_{v\lambda}|
    \right] \\
  S_\beta
  &=\sum_\lambda\left[
    j_{0\lambda}^2+|\ve j_{v\lambda}|^2
    -2s_{j\lambda}j_{0\lambda}|\ve j_{v\lambda}|
    \right]
  \label{eq:expanded-current}
\end{align}
This form preserves the transverse quadratic response
$|\ve j_{v\lambda}|^2$ in the zero-norm limit.

\subsection{Spin-Isotropic Limit}

The signed norms in eqs~\eqref{eq:nc-invariants} and
\eqref{eq:eigen-currents} are not differentiable at zero.  Their
spin-covariant longitudinal continuations are
\begin{align}
  s_\eta|\ve c|&\longrightarrow \ve e\cdot\ve c
  \nonumber\\
  s_\tau|\ve K|&\longrightarrow \ve e\cdot\ve K
  \label{eq:continuations}\\
  s_{j\lambda}|\ve j_{v\lambda}|
  &\longrightarrow \ve e\cdot\ve j_{v\lambda}
  \nonumber
\end{align}
The transverse second-order remainder is evaluated from the corresponding
vector norm before the limit is taken.  As $m\rightarrow0$, the moving spin
frame itself becomes undefined.  The physical kernel nevertheless has a
finite, spin-rotation-invariant limit, obtained directly from the
closed-shell scalar and vector blocks in eq~\eqref{eq:closed-hessian}.
Thus, the Kramers-restricted formulation is also the regular continuation of
the general non-collinear kernel at a spin-isotropic reference.

\section*{Data Availability Statement}
The data underlying this study are openly available in Zenodo
at DOI 10.5281/zenodo.21879659. This includes
\begin{itemize}
    \item Energies and eigenvalues of Au$_2$ (ASCII txt)
    \item Complete NMR results (ods and csv)
    \item Complete EPR results (ods, csv, and ASCII txt)
    \item Complete M\"ossbauer results (ods, csv, and ASCII txt)
    \item TDDFT UV/vis and near-infrared data (ASCII txt)
    \item $GW$-BSE X-ray absorption data (ods and csv)
    \item TDDFT X-ray absorption data (ASCII txt)
    \item Auxiliary basis set for TDDFT/$GW$-BSE (ASCII txt)
    \item Optimized structure of the bismuthinidene molecule (xyz)
    \item SAP-X2C basis sets as Fortran header files (h),
    README files on the construction and assessment of the
    compact basis sets (ASCII txt) as well as a Python script
    for the refit itself with TURBOMOLE (py)
    \item Extended H{\"u}ckel vectors for SCF initial guess of Fr--Og
    in TURBOMOLE format (txt)
\end{itemize}

\section*{Author Contributions and Declarations}
\noindent \textbf{Yannick J. Franzke}: Conceptualization (lead); Data curation (equal);
Formal analysis (equal); Investigation (equal); Methodology (lead);
Software (equal); Validation (equal); Visualization (equal);
Writing – original draft (lead); Writing – review \& editing (equal).
 \\
\noindent \textbf{Christof Holzer}: Conceptualization (supporting); Data curation (equal);
Formal analysis (equal); Investigation (equal); Methodology (supporting);
Software (equal); Validation (equal); Visualization (equal);
Writing – original draft (supporting); Writing – review \& editing (equal).

\begin{acknowledgement}
The authors thank Michal Repisky (Troms{\o}) for sharing a binary of the
ReSpect program to perform four-component calculations (V5.1.0-beta, 2019).
We further thank Michael E.\ Harding (KIT) for providing an auxiliary basis
set for the X2C-TDDFT and $GW$-BSE calculations (cbas) herein. Y.J.F.\ further
thanks Danh X.\ Ngo (UC Berkeley) for discussions on the magnetic and optical
properties of hydrindacene-based organometallic compounds.
\end{acknowledgement}
\medskip

\textbf{Notes} The AI programming tool Codex by OpenAI was used to
convert the SAP basis set of ref~\citenum{Lehtola.Visscher.ea:Efficient.2020}
into a header file for the TURBOMOLE program suite. Refitting of the
SAP basis sets was done with the help of Claude Opus by Anthropic.
Dyall basis sets of the 5f elements and beyond were imported with
Claude Opus. H\"uckel vectors were constructed with this AI tool based
on standard workflows of refs~\citenum{Franzke.Spiske.ea:Segmented.2020,
Pollak.Weigend:Segmented.2017, Pollak:Technische.2017}. Occupations for the
SCF guess of Fr--Og with a superposition of atomic densities were further
straightforwardly generated based on the elements Cs--Rn. New radii data
for the DFT grids \cite{Treutler.Ahlrichs:Efficient.1995, Treutler:Entwicklung.1995}
assume a value of 1.75\,{\AA} for U--Og, as no Bragg--Slater radii are
currently available.

\bibliography{literature}

\end{document}